\RequirePackage{lineno}
\documentclass[twocolumn,showpacs,preprintnumbers,amsmath,amssymb,superscriptaddress]{revtex4}

\newcommand{\nuebar}{\ensuremath{\overline{\nu}_{e}}}

\usepackage{graphicx,bm,epsf,float,amssymb,multirow,dcolumn} 

\begin{document}

\title{PROSPECT - A Precision Reactor Oscillation and Spectrum Experiment at Short Baselines}

\affiliation{Chemistry Department, Brookhaven National Laboratory, Upton, NY 11973}
\affiliation{Physics Department, Brookhaven National Laboratory, Upton, NY 11973}
\affiliation{Department of Physics, Drexel University, Philadelphia, PA 19104-2875}
\affiliation{Nuclear Nonproliferation Division, Idaho National Laboratory, Idaho Falls, ID 83401}
\affiliation{Department of Physics, Illinois Institute of Technology,  Chicago, IL 60616}
\affiliation{Physics Division, Lawrence Berkeley National Laboratory, Berkeley, CA 94720}
\affiliation{Physics Division, Lawrence Livermore National Laboratory, Livermore, CA 94550} 
\affiliation{Department of Physics, Le Moyne College, Syracuse, NY 13214}
\affiliation{National Institute of Standards and Technology, Gaithersburg, MD 20899}
\affiliation{Oak Ridge National Laboratory, Oak Ridge, TN 37831} 
\affiliation{Department of Physics, Temple University, Philadelphia, PA 19122}
\affiliation{Department of Physics, University of Tennessee, Knoxville, TN 37996} 
\affiliation{Center for Neutrino Physics, Virginia Tech, Blacksburg, VA 24061} 
\affiliation{Department of Physics, University of Waterloo, Waterloo, ON N2L 3G1, Canada}
\affiliation{Department of Physics, College of William and Mary, Williamsburg, VA 23187}
\affiliation{Department of Physics, University of Wisconsin, Madison, WI 53706} 
\affiliation{Department of Physics, Yale University, New Haven, CT 06520}

\author{J.~Ashenfelter}
\affiliation{Department of Physics, Yale University, New Haven, CT 06520} 

\author{A.B.~Balantekin}
\affiliation{Department of Physics, University of Wisconsin, Madison, WI 53706} 

\author{H.R.~Band}
\affiliation{Department of Physics, Yale University, New Haven, CT 06520} 

\author{G.~Barclay}
\affiliation{Oak Ridge National Laboratory, Oak Ridge, TN 37831} 

\author{C.~Bass}
\affiliation{Department of Physics, Le Moyne College, Syracuse, NY 13214}


\author{N.S.~Bowden}
\affiliation{Physics Division, Lawrence Livermore National Laboratory, Livermore, CA 94550} 

\author{C.D.~Bryan}
\affiliation{Oak Ridge National Laboratory, Oak Ridge, TN 37831} 

\author{J.J.~Cherwinka} 
\affiliation{Department of Physics, University of Wisconsin, Madison, WI 53706} 

\author{R.~Chu}
\affiliation{Department of Physics, University of Tennessee, Knoxville, TN 37996} 
\affiliation{Oak Ridge National Laboratory, Oak Ridge, TN 37831} 

\author{T.~Classen}
\affiliation{Physics Division, Lawrence Livermore National Laboratory, Livermore, CA 94550} 

\author{D.~Davee}
\affiliation{Department of Physics, College of William and Mary, Williamsburg, VA 23187} 

\author{D.~Dean}
\affiliation{Oak Ridge National Laboratory, Oak Ridge, TN 37831} 

\author{G.~Deichert}
\affiliation{Oak Ridge National Laboratory, Oak Ridge, TN 37831} 

\author{M.~Diwan}
\affiliation{Physics Department, Brookhaven National Laboratory, Upton, NY 11973}

\author{M.~J.~Dolinski}
\affiliation{Department of Physics, Drexel University, Philadelphia, PA 19104-2875}

\author{J.~Dolph}
\affiliation{Physics Department, Brookhaven National Laboratory, Upton, NY 11973}

\author{D.A.~Dwyer}
\affiliation{Physics Division, Lawrence Berkeley National Laboratory, Berkeley, CA 94720}

\author{Y.~Efremenko}
\affiliation{Department of Physics, University of Tennessee, Knoxville, TN 37996} 
\affiliation{Oak Ridge National Laboratory, Oak Ridge, TN 37831} 

\author{S.~Fan}
\affiliation{Department of Physics, University of Tennessee, Knoxville, TN 37996} 
\affiliation{Oak Ridge National Laboratory, Oak Ridge, TN 37831} 

\author{A. Galindo-Uribarri}
\affiliation{Oak Ridge National Laboratory, Oak Ridge, TN 37831} 

\author{K. Gilje}
\affiliation{Department of Physics, Illinois Institute of Technology,  Chicago, IL 60616}

\author{A.~Glenn}
\affiliation{Physics Division, Lawrence Livermore National Laboratory, Livermore, CA 94550} 

\author{M.~Green}
\affiliation{Oak Ridge National Laboratory, Oak Ridge, TN 37831} 

\author{K. Han}
\affiliation{Department of Physics, Yale University, New Haven, CT 06520} 

\author{S.~Hans}
\affiliation{Chemistry Department, Brookhaven National Laboratory, Upton, NY 11973}

\author{K.M.~Heeger}
\affiliation{Department of Physics, Yale University, New Haven, CT 06520} 

\author{B.~Heffron}
\affiliation{Department of Physics, University of Tennessee, Knoxville, TN 37996} 
\affiliation{Oak Ridge National Laboratory, Oak Ridge, TN 37831} 

\author{L.~Hu}
\affiliation{Chemistry Department, Brookhaven National Laboratory, Upton, NY 11973}

\author{P.~Huber}
\affiliation{Center for Neutrino Physics, Virginia Tech, Blacksburg, VA 24061} 

\author{D.~E.~Jaffe}
\affiliation{Physics Department, Brookhaven National Laboratory, Upton, NY 11973}

\author{Y.~Kamyshkov}
\affiliation{Department of Physics, University of Tennessee, Knoxville, TN 37996} 

\author{S.~Kettell}
\affiliation{Physics Department, Brookhaven National Laboratory, Upton, NY 11973}

\author{C.~Lane}
\affiliation{Department of Physics, Drexel University, Philadelphia, PA 19104-2875}

\author{T.J.~Langford}
\affiliation{Department of Physics, Yale University, New Haven, CT 06520} 

\author{B.R.~Littlejohn}
\affiliation{Department of Physics, Illinois Institute of Technology,  Chicago, IL 60616}

\author{D. Martinez}
\affiliation{Department of Physics, Illinois Institute of Technology,  Chicago, IL 60616}

\author{R.D.~McKeown}
\affiliation{Department of Physics, College of William and Mary, Williamsburg, VA 23187} 

\author{M.P.~Mendenhall}
\affiliation{National Institute of Standards and Technology, Gaithersburg, MD 20899}

\author{S.~Morrell}
\affiliation{Nuclear Nonproliferation Division, Idaho National Laboratory, Idaho Falls, ID 83401}

\author{P.~Mueller}
\affiliation{Oak Ridge National Laboratory, Oak Ridge, TN 37831} 

\author{H.P.~Mumm}
\affiliation{National Institute of Standards and Technology, Gaithersburg, MD 20899}

\author{J.~Napolitano}
\affiliation{Department of Physics, Temple University, Philadelphia, PA 19122}

\author{J.S.~Nico}
\affiliation{National Institute of Standards and Technology, Gaithersburg, MD 20899}

\author{D.~Norcini}
\affiliation{Department of Physics, Yale University, New Haven, CT 06520} 

\author{D.~Pushin}
\affiliation{Department of Physics, University of Waterloo, Waterloo, ON N2L 3G1, Canada}

\author{X.~Qian}
\affiliation{Physics Department, Brookhaven National Laboratory, Upton, NY 11973}

\author{E.~Romero}
\affiliation{Department of Physics, University of Tennessee, Knoxville, TN 37996} 
\affiliation{Oak Ridge National Laboratory, Oak Ridge, TN 37831} 

\author{R.~Rosero}
\affiliation{Chemistry Department, Brookhaven National Laboratory, Upton, NY 11973}

\author{B.S.~Seilhan}
\affiliation{Physics Division, Lawrence Livermore National Laboratory, Livermore, CA 94550} 

\author{R.~Sharma}
\affiliation{Physics Department, Brookhaven National Laboratory, Upton, NY 11973}

\author{P.T.~Surukuchi}
\affiliation{Department of Physics, Illinois Institute of Technology,  Chicago, IL 60616}

\author{S.J.~Thompson}
\affiliation{Nuclear Nonproliferation Division, Idaho National Laboratory, Idaho Falls, ID 83401}

\author{R.L.~Varner}
\affiliation{Oak Ridge National Laboratory, Oak Ridge, TN 37831} 

\author{B.~Viren}
\affiliation{Physics Department, Brookhaven National Laboratory, Upton, NY 11973}

\author{W.~Wang}
\affiliation{Department of Physics, College of William and Mary, Williamsburg, VA 23187} 

\author{B.~White}
\affiliation{Oak Ridge National Laboratory, Oak Ridge, TN 37831} 

\author{C.~White}
\affiliation{Department of Physics, Illinois Institute of Technology,  Chicago, IL 60616}

\author{J.~Wilhelmi}
\affiliation{Department of Physics, Temple University, Philadelphia, PA 19122}

\author{C.~Williams}
\affiliation{Oak Ridge National Laboratory, Oak Ridge, TN 37831} 

\author{R.E.~Williams}
\affiliation{National Institute of Standards and Technology, Gaithersburg, MD 20899}

\author{T.~Wise}
\affiliation{Department of Physics, Yale University, New Haven, CT 06520} 

\author{H. Yao}
\affiliation{Department of Physics, College of William and Mary, Williamsburg, VA 23187} 

\author{M.~Yeh}
\affiliation{Chemistry Department, Brookhaven National Laboratory, Upton, NY 11973}

\author{N.~Zaitseva}
\affiliation{Physics Division, Lawrence Livermore National Laboratory, Livermore, CA 94550} 

\author{C.~Zhang}
\affiliation{Physics Department, Brookhaven National Laboratory, Upton, NY 11973}

\author{X.~Zhang}
\affiliation{Department of Physics, Illinois Institute of Technology,  Chicago, IL 60616}

\collaboration{The PROSPECT Collaboration}

\date{\today}

\begin{abstract}
Current models of antineutrino production in nuclear reactors predict detection rates and spectra at odds with the existing body of direct reactor antineutrino measurements.  High-resolution antineutrino detectors operated close to compact research reactor cores can produce new precision measurements useful in testing explanations for these observed discrepancies involving underlying nuclear or new physics.  Absolute measurement of the $^{235}$U-produced antineutrino spectrum can provide additional constraints for evaluating the accuracy of current and future reactor models, while relative measurements of spectral distortion between differing baselines can be used to search for oscillations arising from the existence of eV-scale sterile neutrinos.  Such a measurement can be performed in the United States at several highly-enriched uranium fueled research reactors using near-surface segmented liquid scintillator detectors.  We describe here the conceptual design and physics potential of the PROSPECT experiment, a U.S.-based, multi-phase experiment with reactor-detector baselines of 7-20 meters capable of addressing these and other physics and detector development goals.  Current R\&D status and future plans for PROSPECT detector deployment and data-taking at the High Flux Isotope Reactor at Oak Ridge National Laboratory will be discussed.


\end{abstract}

\maketitle

\section{Introduction}
\label{sec:intro}

Reactor antineutrino experiments have played an important role throughout the history of neutrino physics and led to many of the key discoveries in the field.  The existence of the neutrino was first experimentally established by observing reactor antineutrinos at Savannah River~\cite{savannah:1956}.  More recently, the KamLAND, Daya Bay, RENO, and Double Chooz experiments have observed disappearance of reactor antineutrinos and provided measurements of the oscillation parameters $\theta_{12}$, $\theta_{13}$, $\Delta m^2_{21}$ and $\Delta m^2_{31}$\,\cite{kamland:2008, An:2012eh, dayabay:2013, reno:2012, doublechooz:2012}.  Together with atmospheric, solar and accelerator experiments, these reactor measurements provide a coherent picture of neutrino propagation and mixing between the three Standard Model neutrino flavors.  A similarly coherent picture of antineutrino \textit{production} in reactors requires precision measurements of the reactor antineutrino energy spectrum.  By making a high-resolution spectrum measurement at short distances from a compact research reactor core, the PROSPECT (Precision Reactor Oscillation and SPECTrum) experiment can provide new inputs valuable to clarifying production models while also addressing unresolved non-standard neutrino mixing hypotheses.

Anomalous results from recent measurements and theoretical calculations do not fit the existing framework of production and oscillation of reactor antineutrinos.   Improved reactor antineutrino flux predictions~\cite{Mueller:2011nm,Huber:2011wv} have resulted in an increase in the predicted interaction rate of $\sim3.5\%$\ with respect to previous calculations~\cite{Schreckenbach:1985ep, Hahn:1989zr, Vogel:1989}.  When combined with experimental data at baselines between 10-100\,m these recent calculations suggest a $\sim$5.6\% difference between the measured and predicted reactor antineutrino flux~\cite{Mention:2011rk, Zhang:2013, dayabay:2014Neutrino}.  Furthermore, new measurements of the absolute reactor antineutrino spectrum from reactor $\theta_{13}$ experiments collectively indicate discrepancies with respect to both the historical and improved reactor flux and spectrum predictions, particularly at antineutrino energies of 5-7~MeV~\cite{reno:2014, doublechooz:2014, dayabay:2014Neutrino}.  These flux and spectrum anomalies can be interpreted as either a sign of new neutrino physics or as-yet-undetermined imperfections in predictions of reactor antineutrino emissions.

If existing reactor flux predictions do not properly take into account the underlying nuclear physics and all relevant systematic uncertainties, confidence levels of reactor-related anomalies may be over-stated.  In this case, new neutrino physics need not be introduced to explain all existing reactor data.  One such effect is the handling of beta branch forbiddenness in beta spectrum conversions: by varying shape factors of utilized virtual beta branches, interaction rate variations of 5\% can be produced along with significant spectral deformations~\cite{Hayes:2013wra}.  
The accuracy of existing beta spectrum measurements has also been questioned~\cite{Dwyer:2014}: if reactor beta and $\nuebar$ spectra are calculated exclusively using existing nuclear data, the result appears to not match existing beta spectrum measurements, but does reproduce the spectral shape 
of recent $\nuebar$ measurements.  
Additional measurements of fine structure in the $\nuebar$ spectrum and of spectra from different combinations of fission isotopes are necessary to test these hypotheses and provide constraints on future calculations.  This motivates high-resolution measurement of the absolute $\nuebar$ spectrum with both highly-enriched (HEU) and conventional low-enriched (LEU) reactor fuel.

Elucidation of the true reactor antineutrino flux and spectrum is also valuable in other  oscillation physics and nuclear applications contexts.  While new relative measurements of $\theta_{13}$ and $\Delta m^2_{31}$ are unlikely to be significantly biased by imperfect flux predictions, confidence in oscillation results from absolute measurements, such as those proposed at future reactor mass hierarchy experiments, could be improved with higher confidence in reactor flux prediction.  In nuclear safeguards applications, to reliably determine a core's plutonium content in real time for non-proliferation purposes, as illustrated in~\cite{HuberKorea, HuberIran}, standard measurements of Pu-free (HEU) and Pu-rich (LEU) cores are significantly more useful than inaccurate reactor flux models of HEU and LEU cores.  In the nuclear applications community, measurement of the individual contributors to the reactor $\nuebar$ flux in the 5-7 MeV region may provide valuable cross-checks to gamma spectroscopy in determining the properties of major beta decay branches in newly discharged nuclear fuel~\cite{DecayHeat:2007}.  These broader implications further motivate the need for HEU and LEU absolute spectrum measurements with high energy resolution detectors.

By making $\nuebar$ energy spectrum measurements at a variety of short reactor-detector baselines, a detector measuring the absolute reactor flux and spectrum can simultaneously perform sensitive searches for new neutrino physics suggested by the reactor flux anomaly.  In particular, it has been suggested that the reactor flux deficit may be the signature of additional sterile neutrino states with mass splittings of the order of $\sim1~{\rm eV}^2$~\cite{Abazajian:2012ys, Giunti:2011hn}. Additional sterile neutrino mass states with $\Delta m^2 \sim 1$\, eV$^2$ beyond the 3 active neutrinos would yield an oscillation effect for reactor $\nuebar$ traveling over meter-long baselines.  This hypothesis can also explain other anomalous neutrino measurements, such as event excesses in $\nuebar$ and $\nu_e$ appearance channels ~\cite{lsnd:2001, miniboone:2012} and rate deficits observed in solar neutrino detector calibrations with high-intensity $\nu_e$ sources~\cite{Giunti:2011hn}.  However, a variety of results from $\nu_{\mu}$ disappearance measurements show no evidence for non-standard behavior in a similar region of oscillation phase space~\cite{Kopp:2013}.  Furthermore, the presence of sterile-mediated oscillations is unlikely to resolve the discrepancy between measured and predicted reactor spectra, which does not exhibit the requisite $L/E$ behavior~\cite{dayabay:2014sterile}.


A relative measurement of $L/E$ oscillations at short reactor baselines can be used to perform a definitive search for oscillations in the parameter space suggested by global fits to existing anomalies~\cite{Abazajian:2012ys}.  Existing km-scale reactor experiments, while highly precise, cannot definitively probe oscillation lengths of this order~\cite{Abe:2011fz, An:2012eh, reno:2012}: at these baselines any oscillation effect due to potential sterile states is averaged by finite detector resolution to yield an effective rate deficit.  Instead, a new experiment at very short baselines in a controlled research environment is needed to fully disentangle reactor flux and spectrum uncertainties from possible sterile neutrino oscillations and other effects.  However, at these short baselines, a detector's position resolution, energy resolution, and the finite dimensions of the reactor core become important. This motivates the use of segmented detectors in close proximity to reactors with compact cores of less than $\mathcal{O}$(1~m) dimensions~\cite{VSBL, MultSBL}.  In the United States, these central experimental criteria can be met at several HEU research reactor facilities.  These facilities possess other advantageous features for an oscillation measurement that will be discussed further in Section~\ref{sec:researchRx}.

The experimental challenges involved in making a precise spectral measurement at short distances from a reactor appear tractable based on the recent experience of other reactor experiments and R\&D efforts.  The precise relative energy calibration and control of detector response systematics necessary for a relative oscillation measurement at multiple baselines have been well-demonstrated in recent successful multi-detector $\nuebar$ experiments~\cite{DYB:2013cpc, dayabay:2013, reno:2012}.  The ability to reduce and reject inverse beta decay backgrounds in the absence of significant overburden is extremely challenging, but efforts incorporating particle identification techniques and optimized shielding designs suggest a path forward~\cite{Abbes:1995nc,Kiff:2011,Nelson:2011ux,Bowden:2012um, panda:2012, Belov:2013qwa}.  Careful detector and shielding designs will be required to address these challenges, and can be validated by focused demonstrations at host reactor sites. 

The National Institute of Standards and Technology (NIST) \cite{NIST}, Oak Ridge National Laboratory (ORNL) \cite{HFIR}, and  Idaho National Laboratory (INL) \cite{ATR}  operate powerful, highly compact research reactors and have identified potential sites for the deployment of compact $\nuebar$ detectors at distances between 4-20~m  from the reactor cores.  By deploying segmented liquid scintillator antineutrino detectors at any of these three US-based reactor sites, the PROSPECT experiment offers a unique opportunity to provide the first-ever high-resolution measurement of the HEU $\nuebar$ spectrum while searching for $\nuebar$ oscillations at very short baselines.  The PROSPECT absolute spectrum measurement will be complimentary to new spectral measurements by $\theta_{13}$ experiments in providing new inputs for reactor flux modeling valuable the oscillation and applications communities.  In addition, a high confidence-level discovery of sterile neutrinos is possible with 1 (3) years of data taking.  The PROSPECT collaboration has begun deployment of a series of prototype detectors at the High Flux Isotope Reactor (HFIR) at ORNL.   These will provide verification of supporting detection techniques and simulations in preparation for installation of one or two multi-ton-scale optically segmented Li-loaded liquid scintillator detectors capable of discriminating nuclear from electromagnetic interactions based on scintillation time profiles.

This note describes the PROSPECT experiment and its potential to produce a highly precise measurement of the $\nuebar$ spectrum, definitely resolve one of the outstanding anomalies in neutrino physics, and further develop the application of advanced scintillators and near-surface detectors to reactor monitoring and safeguards.  Section II will describe the advantageous features of research reactors as sites for precision oscillation experiments.  Section III will outline the experimental strategy and physics potential of the PROSPECT experiment.  Sections IV, V, and VI will present in more detail the potential reactor sites, proposed detector designs, and expected backgrounds, respectively.  Finally, Section VII will provide an overview of current R\&D activities.

\section{Research Reactors as Laboratories for Precision Studies}
\label{sec:researchRx}

The large antineutrino flux produced by nuclear power reactors has led to such sites being a preferred venue for reactor neutrino studies over the past two decades. However, research reactors operated by scientific organizations and national laboratories possess many advantages for precision neutrino physics studies, especially at short baselines. While research reactors operate at  lower powers than commercial plants, it is often possible to gain access to locations closer to the reactor core, partially compensating the reduction in flux. In addition, the compact core geometry, core composition, and operations of research reactors offer unique advantages for a reactor experiment at very short baselines.

The primary feature of research reactors relevant to precision studies of short-baseline oscillations with a length scale of $\mathcal{O}$(m) is the core geometry and composition. While research reactor core geometries can vary considerably depending upon their intended application, their spatial extent is of the order of 1~m, less than that of all existing power reactors, and less than that of the suggested sterile-mediated neutrino oscillation wavelength.  Additionally, many research reactors use fuel that is comprised primarily of Highly Enriched Uranium (HEU). Unlike the Low Enriched Uranium (LEU) or Mixed Oxide (MOX) fuel used at power reactors, there is insufficient $^{238}$U present in research reactor fuel to breed substantial amounts of Pu, in particular $^{239}$Pu. Accordingly, essentially all antineutrinos emitted by HEU fueled research reactors derive from  $^{235}$U daughters, and the core fission fractions are constant throughout a reactor operational cycle. This is in contrast to the behavior of power reactors, where Pu breeding results in time-varying  power contributions from $^{235}$U, $^{239}$Pu, and $^{241}$Pu, and therefore time variation in the emitted antineutrino flux and spectrum. The near static character of research reactor antineutrino emissions is advantageous for precision studies since it substantially reduces the importance of complicated reactor evolution codes to predict fission fractions throughout the reactor cycle.

Unlike power reactors, research reactors operate frequent short cycles. The resulting reactor off periods provide important opportunities for background characterization. Since research reactor duty cycles are typically no greater than 70\%, there is a substantial period of reactor outage time during which to obtain direct measurement of background at such facilities.  During reactor-off periods, it may also be possible to observe antineutrinos emitted by long lived isotopes from the research reactors' spent nuclear fuel.

\begin{figure*}[htb!]
\centering
\includegraphics*[trim=0.1cm 0.1cm 0.1cm 0.1cm, clip=true, width=0.8\textwidth]{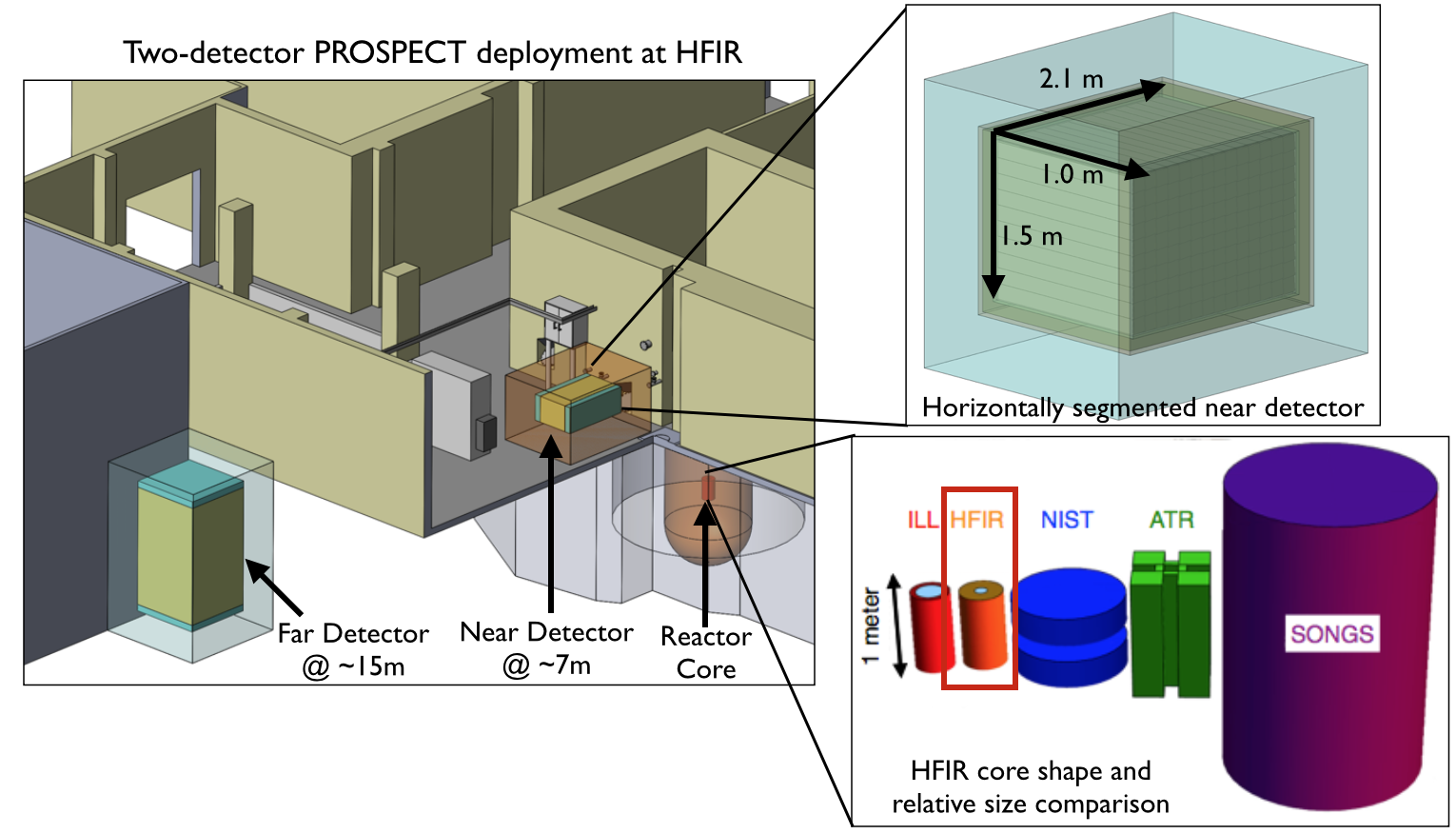}
\caption{An example rendering of the PROSPECT experimental deployment at the HFIR reactor site. The Phase~I detector is moveable, and can be placed at a closest distance of $\sim$7~m to the reactor core center, while the larger Phase~II detector can be placed at closest distance of $\sim$15~m to the core center.}
\label{fig:expt}
\end{figure*}

Research reactors typically maintain detailed neutronic core models that are used to predict neutron fluxes and power densities cycle-by-cycle. This is often important as irradiation experiments that are exchanged between cycles can have a large local effect on these parameters. These models and their outputs are typically available to all users of the facility.  This is potentially important  for a short-baseline reactor experiment, as the core power, and hence baseline, distribution may vary slightly cycle to cycle. The ease with which this important information can be accessed is in contrast to the situation at commercial plants where the core models are typically proprietary, and special arrangements must be made with the plant operator and/or fuel vendor.

\section{Experimental Strategy and Physics Reach}
\label{sec:expt_reach}

The PROSPECT experiment will utilize the Inverse Beta Decay (IBD) reaction $\overline{\nu}_e +p \rightarrow e^+ + n$ with a threshold of 1.8 MeV to measure the flux and energy spectrum of reactor $\overline{\nu}_e$.  Interactions will take place in and around the experiment's proton-rich liquid scintillator target, producing a pair of triggers correlated in space and time.  Prompt signal triggers produced by the kinetic energy deposition and annihilation of an IBD positron will mirror the energy spectrum of the incident reactor antineutrinos from $\sim$1-8 MeV and display a scintillation time profile characteristic of an electromagnetic interaction.  Delayed triggers produced by IBD neutron capture on lithium and subsequent decay to a high-energy $^4$He and $^3$H will exhibit a highly spatially-localized and quenched energy deposition of approximately 0.5~MeV$_{ee}$ with a scintillation time profile characteristic of a nuclear interaction.   By selecting only spatially and time-coincident triggers with the appropriate prompt and delayed energy, timing, and spatial profiles, the PROSPECT detector will select antineutrino interaction candidates with high detection efficiency and good energy resolution.  Further rejection of cosmic backgrounds can be achieved if necessary by incorporating an external muon veto system or by utilizing fiducialization of the target volume.

The full PROSPECT experiment consists of an $\mathcal{O}$(1\,m$^3$)-sized near detector target at distances of $\sim$7~m to $\sim$10.5~m and an $\mathcal{O}$(10\,m$^3$)-sized far detector target at distances of 15-20\,m.  These detectors will consist of optically separated functionally identical sub-volumes that provide precise, reliable position  and energy resolution, consistent spectral response, and uniform background rejection capabilities.  Preliminary detector and reactor parameters are listed in Table~\ref{tab:DefaultChar}.  A rendering of how the experiment can be configured at the HFIR reactor site is shown in Fig.~\ref{fig:expt}.  The technical design proposed to meet these physics requirements is described in further detail in Section~\ref{sec:detector}.  The PROSPECT experiment's experimental arrangement provides excellent sensitivity to neutrino oscillations over a broad range of mass splittings, and can provide precision absolute spectral measurements of an HEU reactor core. 

\begin{table}[!htb]{%
    \begin{tabular}{l | l | l } \hline
      \multicolumn{2}{c|}{Parameter} & Value\\ \hline
      \multirow{4}{*}{Reactor} & Power & 85 MW \\
      & Shape & cylindrical  \\
      & Radius & 0.2~m \\
      & Height & 0.3~m \\
      & Fuel & HEU \\
      & Duty cycle & 41\% reactor-on \\ \hline
      \multirow{8}{*}{Detector} & Cross-section (near) & 1.0~m$\times$1.5~m \\
      & Cross-section (far) & 1.0~m$\times$3.0~m \\
      & Baseline coverage (near) & 2.1~m \\
      & Baseline coverage (far) & 4.2~m \\
      & Efficiency & 30\%  \\  
      & Proton density & 5.5$\times$10$^{28}\frac{p}{m^3}$ \\
      & Position resolution & 15~cm   \\
      & Energy resolution & 4.5\%/$\sqrt{E}$ \\ \hline
      \multirow{2}{*}{Background}& S:B ratio & 1 \\
      & Background shape & 1/E$^2$ + Flat \\ \hline
      \multirow{3}{*}{Other} & Run Time & 1 or 3 calendar years  \\
      & Closest distance (near) & $\sim$7~m \\ 
      & Closest distance (far) & $\sim$15~m \\ \hline
    \end{tabular}}
  \caption{Nominal detector and reactor parameters for the proposed Phase~I experiment, in the case of deployment at HFIR.  The Phase~II parameters are identical with the exception of the far detector placement and size.}
  \label{tab:DefaultChar}
\end{table}

The two-detector arrangement of the proposed experiment allows staging in two phases.  Phase~I consists of measurement with the near detector for one to three calendar years, and, as described below, provides a high-statistics absolute spectral measurement with better than 3$\sigma$ sensitivity to a broad range of oscillation parameters.  The Phase~I detector will be moveable to allow for deployment at a 'front' location described in Table~\ref{tab:DefaultChar}, as well as a 'back' location at a horizontal reactor-detector baseline increased by 1.5~m, giving a total Phase I baseline coverage of $\sim$7-11~m.  This ability increases the oscillation sensitivity of the Phase I detector, and will allow for valuable cross-checks on any positive oscillation signal.  Furthermore, a moveable detector can be used to map out spatial variations in detector backgrounds during reactor-off periods.

Phase~II consists of a three year run of both detectors, which provides a significantly increased statistical sample for an absolute spectral measurement while extending the region of sensitivity and providing a conclusive test of most of the suggested oscillation parameter space.  By utilizing a phased approach, PROSPECT will be able to build on the experience and knowledge gained during deployment and data-taking with the PROSPECT prototype and Phase I detectors and optimize the Phase II detector for the signature or parameter space of interest.





\subsection{Measurement of the Reactor Antineutrino Spectrum}

PROSPECT intends to measure the energy spectrum of $\overline{\nu}_e$
emitted by a highly-enriched uranium (HEU) nuclear reactor to a
precision exceeding that provided by current best predictions.
Three specific
features of this measurement serve to strongly constrain reactor
models: the general HEU spectral shape, high-resolution spectral
features, and the deviation from recent LEU spectral measurements.

\begin{figure}[htb!]
\includegraphics[trim=0.3cm 0.1cm 1.7cm 0.3cm, clip=true, width=0.99\columnwidth]{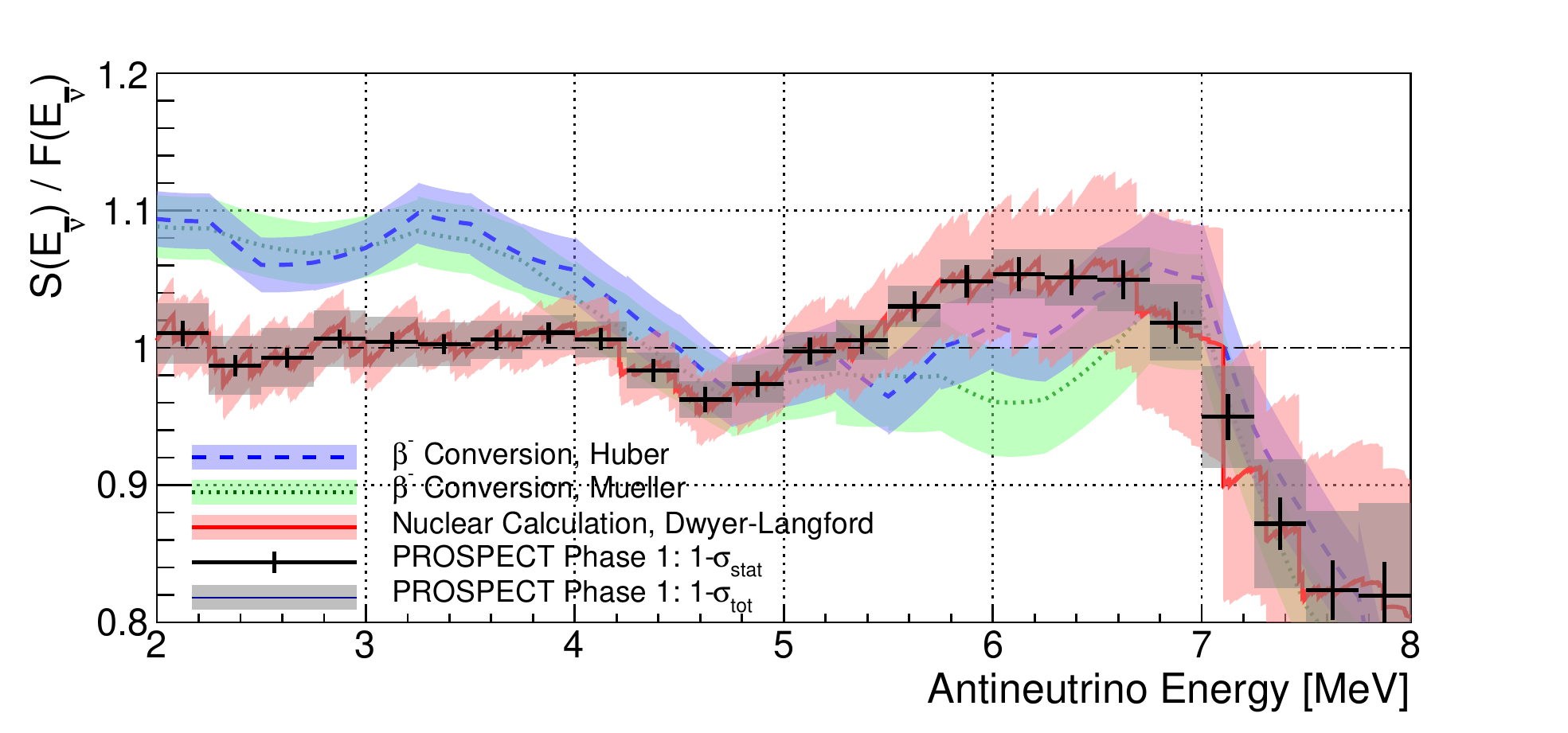}
\includegraphics[trim=0.3cm 0.1cm 1.7cm 0.3cm, clip=true, width=0.99\columnwidth]{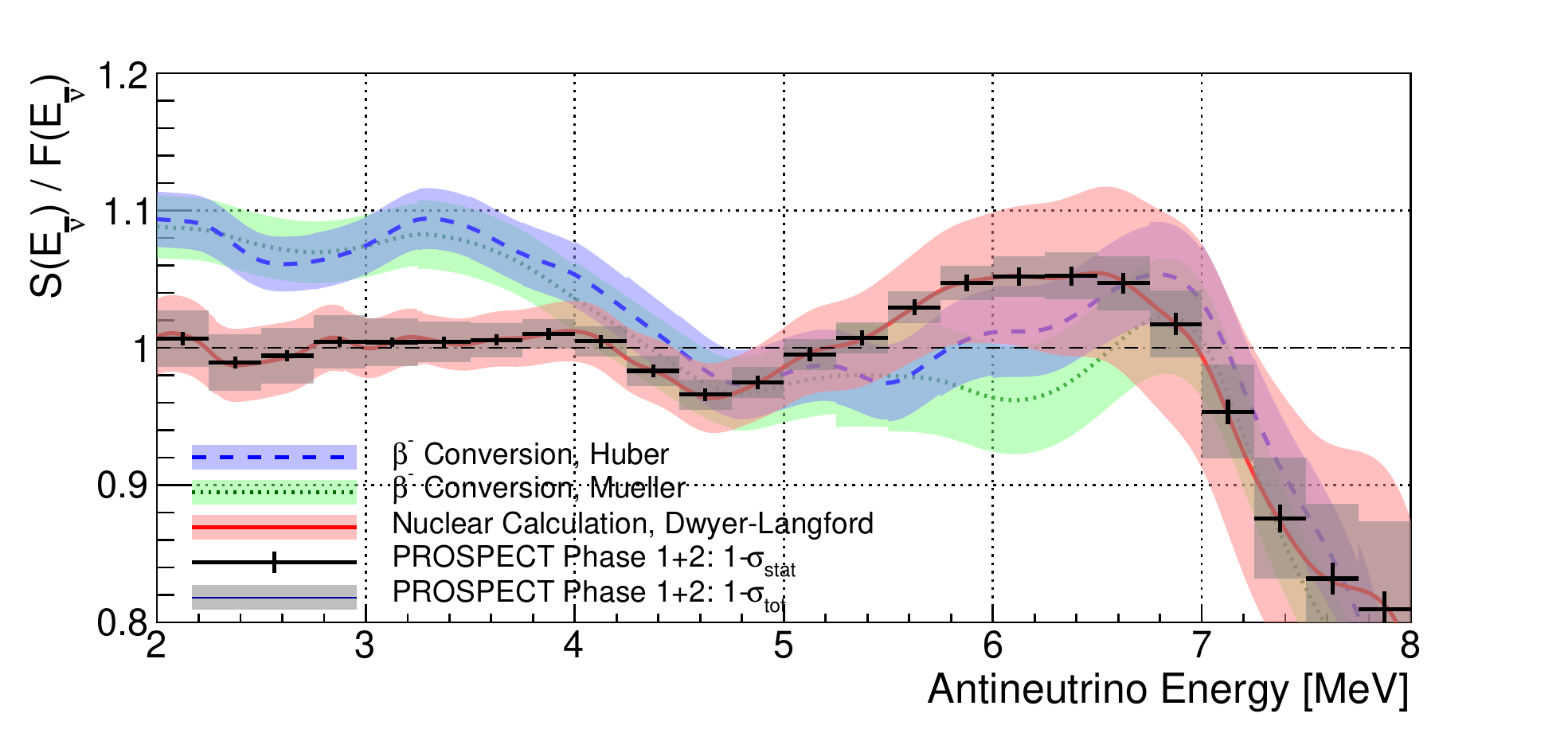}
\caption{ {\em Upper:} Three models of the energy spectrum of
  $\overline{\nu}_e$ emitted by fission daughters of $^{235}$U.  To
  highlight the shape differences between models, they are shown in
  ratio to a smooth approximation of the spectrum,
  $F(E_{\overline{\nu}})$.  The $1\sigma$ statistical precision of
  the fiducialized Phase I (top) and fiducialized Phase II (bottom) measurement (black bars) is included for comparison,
  along with reported $1\sigma$ systematic uncertainties from~\cite{dayabay:2014Neutrino}.
  Total combined unceratinty is indicated by the gray band.
  {\em Lower:} Including a nominal 4.5\%/$\sqrt{E}$ detector
  resolution reduces much of the detailed bin-to-bin fluctuations
  predicted by the nuclear calculation.
\label{fig:specModelComparison}}
\end{figure}

{\bf General HEU Spectrum}: A precise measurement of the antineutrino
energy spectrum from an HEU reactor can strongly discriminate between
existing reactor $\overline{\nu}_e$ flux models.  Fig.~\ref{fig:specModelComparison} shows the
differences between three current models: two based on the
$\beta^-$-conversion method, and one based on {\em ab-initio}
calculation.  To highlight the shape differences between models, they
are shown in ratio to a smooth approximation from~\cite{Vogel:1989}.
The expected PROSPECT
Phase I and II statistical precision is included for reference, along with
spectral systematic uncertainty bands reported by Daya Bay~\cite{dayabay:2014Neutrino}.
PROSPECT  be able to easily discriminate between these models, as well as
determine the spectrum more precisely than any of the predictions.  As demonstrated
in Figure~\ref{fig:specModelComparison}, in
order to make this precision measurement, it is critical to control
systematic uncertainties from backgrounds and detector response.  Simulation
studies and background surveys are currently underway to determine
the spectrum uncertainty contribution from these sources for PROSPECT.

{\bf High-Resolution Spectral Features}: {\em Ab-initio} nuclear
calculations predict significant discontinuities in the
$\overline{\nu}_e$ energy spectrum, as shown in the upper panel of
Fig.~\ref{fig:specModelComparison}. Each discontinuity is caused by
the Coulomb correction to a single beta decay occurring in the
reactor.  A high-resolution measurement of the energy spectrum could
in principle reveal most of the significant decays contributing to the
energy spectrum.  Such {\em reactor spectroscopy} would provide a new
regime for evaluating reactor $\nuebar$ flux models.  Considering a nominal
4.5\%/$\sqrt{E}$ detector resolution for the Phase-I measurement,
the most prominent bin-to-bin fluctuations should still be
identifiable.  Optimizing detector resolution and control of
backgrounds with discontinuous energy spectra are critical to this
measurement.  Studies are currently underway to determine the 
precision of spectroscopic measurements for a variety of
fission isotopes and detector resolutions.  
A Phase-II
measurement would allow for a higher-statistics sample
to further probe any observed structure measured during Phase I.

{\bf Deviation from LEU Measurements}: Models predict a $\sim$10\%
difference in the flux and spectra between LEU and HEU reactor
$\overline{\nu}_e$ emission.  These differences are primarily
attributable to fission of $^{239}$Pu.  Measurement of the energy
spectrum from fission of $^{235}$U at an HEU reactor can be combined
with existing spectral measurements at LEU reactors to extract the
non-$^{235}$U contribution to the spectrum.  Since both measurements
are expected to have $\sim$1\% precision, the differences should be
prominent and provide another route to evaluate and refine reactor
models.

\subsection{Sensitivity to Short-Baseline Oscillation}

By comparing detected IBD positron spectra at differing reactor-detector baselines, short-baseline reactor experiments can have significant sensitivity to neutrino oscillations.  This can be experimentally accomplished by comparing spectra from different portions of a single detector with sufficiently precise position resolution and uniform response or by comparing spectra between separate detector deployments at differing baselines.  Phase I of the PROSPECT experiment will use the former method with a single radially-extended detector to provide sensitivity to mass splittings of order 1-10\,eV$^2$ and address the current best-fit sterile oscillation parameter space at high confidence level.  Phase II of the PROSPECT experiment will use both methods with two detectors to allow precise oscillation measurements below 1\,eV$^2$ and the ability to observe multiple $L/E$ oscillation periods. 


The PROSPECT experiment's sensitivity to neutrino oscillations is evaluated by comparing the detected inverse beta decay prompt events $T_{ij}$ in energy bin $i$ and position bin $j$ to the expected events $M_{ij}$ in the absence of neutrino oscillations and in the presence of a background $B_{ij}$.  For the purposes of these calculations, $T_{ij}$ is taken as $B_{ij}$ plus an oscillated version of $M_{ij}$.  A $\chi^2$ is used to test the hypothesis of no-oscillation and for oscillation parameter estimation in the case of either one or two additional sterile neutrino states, identically to that presented in Ref~\cite{VSBL, MultSBL}:
\begin{eqnarray}
\label{eq:chi2}
\displaystyle
\chi^2 & = & \sum_{i,j}\frac{\left[M_{ij} - (\alpha + \alpha_e^i + \alpha_r^j)T_{ij} - (1+\alpha_b)B_{ij} \right]^2 }{T_{ij}+B_{ij}+\sigma_{b2b}^2(T_{ij}+B_{ij})^2}  \nonumber \\
& + & \displaystyle \frac{\alpha^2}{\sigma^2} + \displaystyle\sum_{j}\left(\frac{\alpha_r^j}{\sigma_r}\right)^2 + \displaystyle\sum_{i}\left(\frac{\alpha_e^i}{\sigma_e^i}\right)^2 + \frac{\alpha_b^2}{\sigma_b^2}.
\end{eqnarray}

The $\chi^2$ sum is minimized with respect to the relevant oscillation parameters and to the nuisance parameters \{$\alpha$, $\alpha_r^j$, $\alpha_e^i$, $\alpha_b$\} characterizing the systematic uncertainties of the measurement, as described in~\cite{PDG}.  These nuisance parameters represent the overall normalization, uncorrelated position normalization, uncorrelated energy spectrum, and background systematics. Associated bounding uncertainties of these systematics are \{$\sigma$, $\sigma_r$, $\sigma_e$, $\sigma_b$\} = \{100\%, 1.0\%, 10\%, 100\%\}.  Reactor flux shape and normalization uncertainties are inflated to ensure that sterile sensitivity arises purely from relative oscillometric effects, and is not improved by assumptions about the absolute reactor flux or spectrum.  An additional uncertainty $\sigma_{b2b}$ of 0.5\% is added to the $\chi^2$ to account for possible background and detector response uncertainties uncorrelated in both position and energy, which are currently under detailed investigation by the collaboration.

The 3+1 neutrino model with one additional sterile neutrino state and a mass splitting of~$\sim$1~eV$^2$ mass is frequently used in the literature to benchmark the sensitivity of new experiments to short-baseline oscillations \cite{Abazajian:2012ys}.  In keeping with this convention, we present the PROSPECT experiment's sensitivity to 3+1 neutrino oscillations for one and two detectors. The short-baseline $\overline{\nu}_e$ survival probability associated with this oscillation is described by
\begin{eqnarray}
  \label{eq:Eq31}
  P_{ee} & = 1-4|U_{e4}|^2(1-|U_{e4}|^2)\sin^2\frac{\Delta m^2 L}{4E} \nonumber \\
  & = 1 - \sin^22\theta_{ee}\sin^2\frac{\Delta m^2 L}{4E},
\end{eqnarray}
with the oscillation amplitude $\sin^22\theta_{ee}$ = $\sin^22\theta_{14}$ = 4$|U_{e4}|^2(1-|U_{e4}|^2)$.

\begin{figure}[htb!pb]
\centering
\includegraphics[trim=0.4cm 0.1cm 0.7cm 0.1cm, clip=true, width=0.47\textwidth]{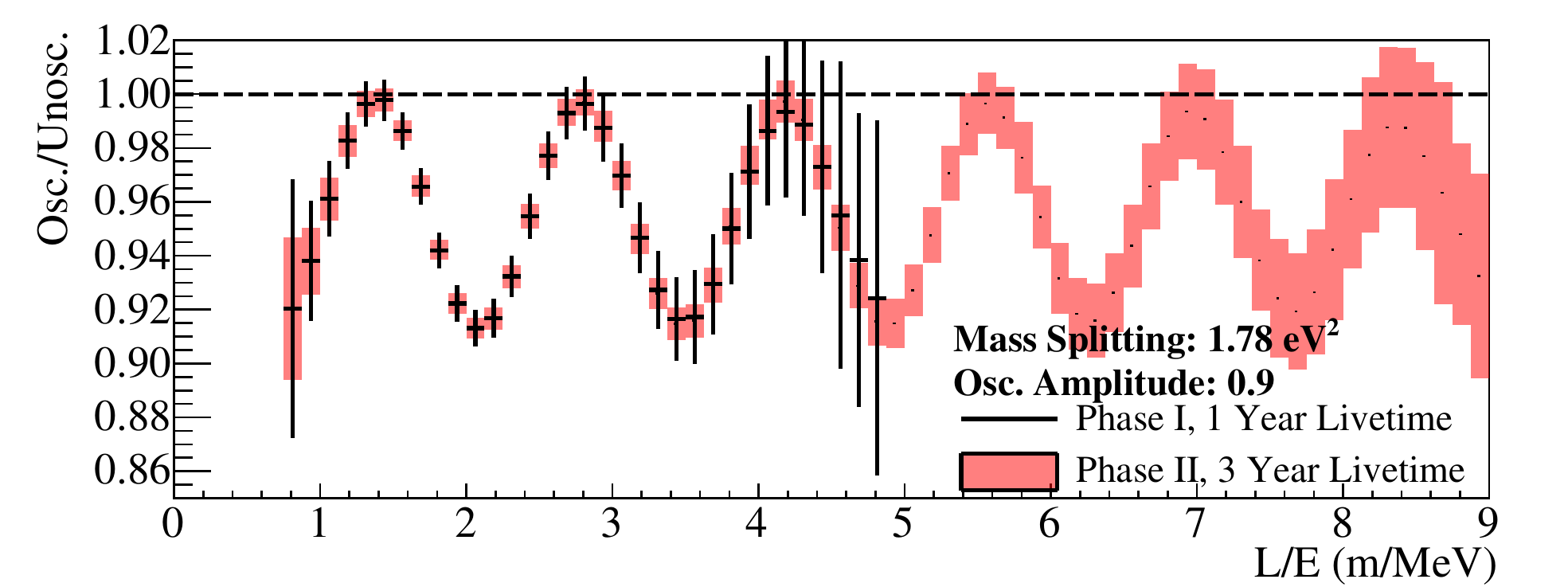}
\includegraphics[trim=0.4cm 0.1cm 0.7cm 0.1cm, clip=true, width=0.47\textwidth]{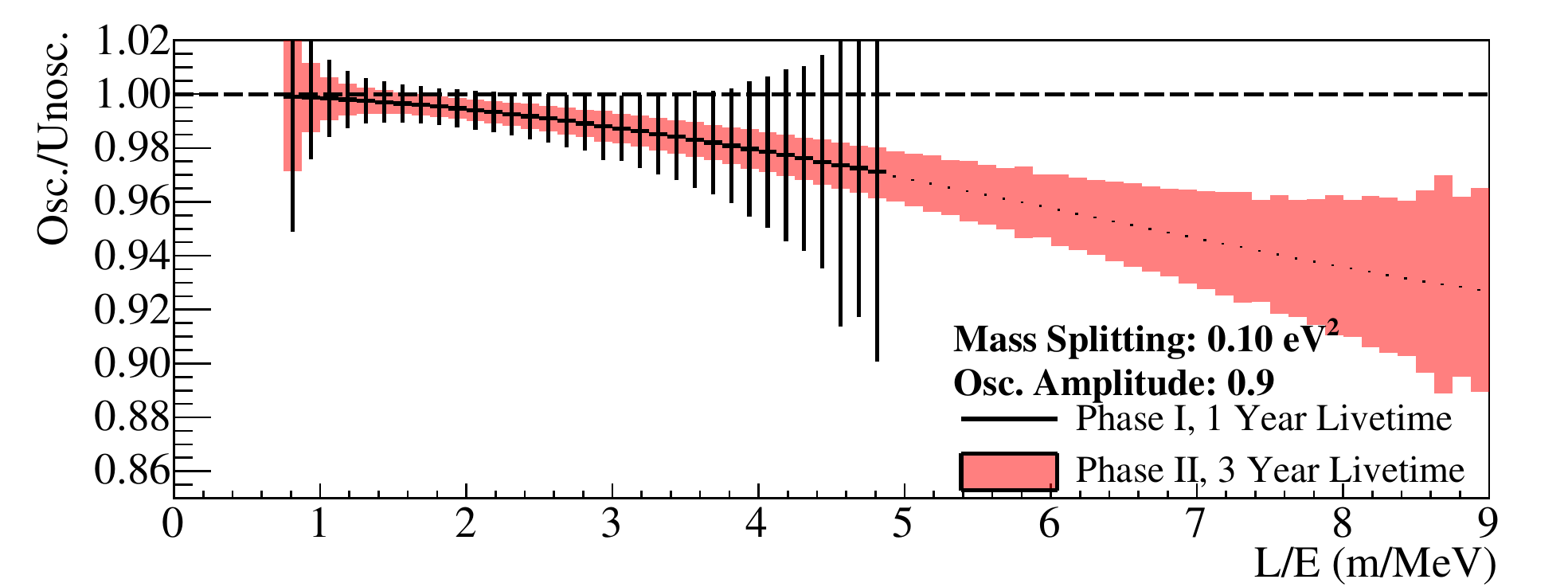}
\caption[]{Oscillated $L/E$ distributions for 3+1 neutrino mixing assuming the 3+1 best-fit oscillation parameters (sin$^2$2$\theta_{13}$,$\Delta$m$^2$) = (0.09, 1.78 eV$^2$) from~\cite{Kopp:2013} (top), and for mixing at lower frequency, (sin$^2$2$\theta_{13}$,$\Delta$m$^2$) = (0.09, 1.78 eV$^2$) (bottom). Error bars are shown with 1 year of data from PROSPECT Phase~I and 3 years of data from the full Phase~II. The default experimental parameters described in Table~\ref{tab:DefaultChar} are used.  In the Phase~II case, the far detector active target mass is 4$\times$ that of the near detector.  Error bars display statistical uncertainties only. }
\label{fig:Osc31}
\end{figure}

Fig.~\ref{fig:Osc31} shows oscillated $L/E$ distributions assuming the existence of one sterile neutrino state for Phase~I and Phase~II of the experiment at two values of $\Delta m^2$.  The measured $L/E$ distributions include smearing from finite core size and from the detector position and energy resolutions shown in Table~\ref{tab:DefaultChar}.  One can see that with a single detector running for one calendar year, the characteristic $L/E$ behavior in the vicinity of the current sterile best-fit region can be clearly mapped out in a manner comparable to previous reactor measurements of $\theta_{13}$ and $\theta_{12}$~\cite{dayabay:2013, kamland:2008}.  As a second detector is added in  Phase~II, the range of $L/E$ coverage increases from 2~m/MeV to greater than 6~m/MeV, allowing for probing of a of wider frequency range and observation of multiple oscillation periods at higher $\Delta m^2$.

\begin{figure}[tb]
\centering
\includegraphics[width=0.48\textwidth, trim=0.3cm 0.1cm 2.5cm 0.1cm, clip=true]{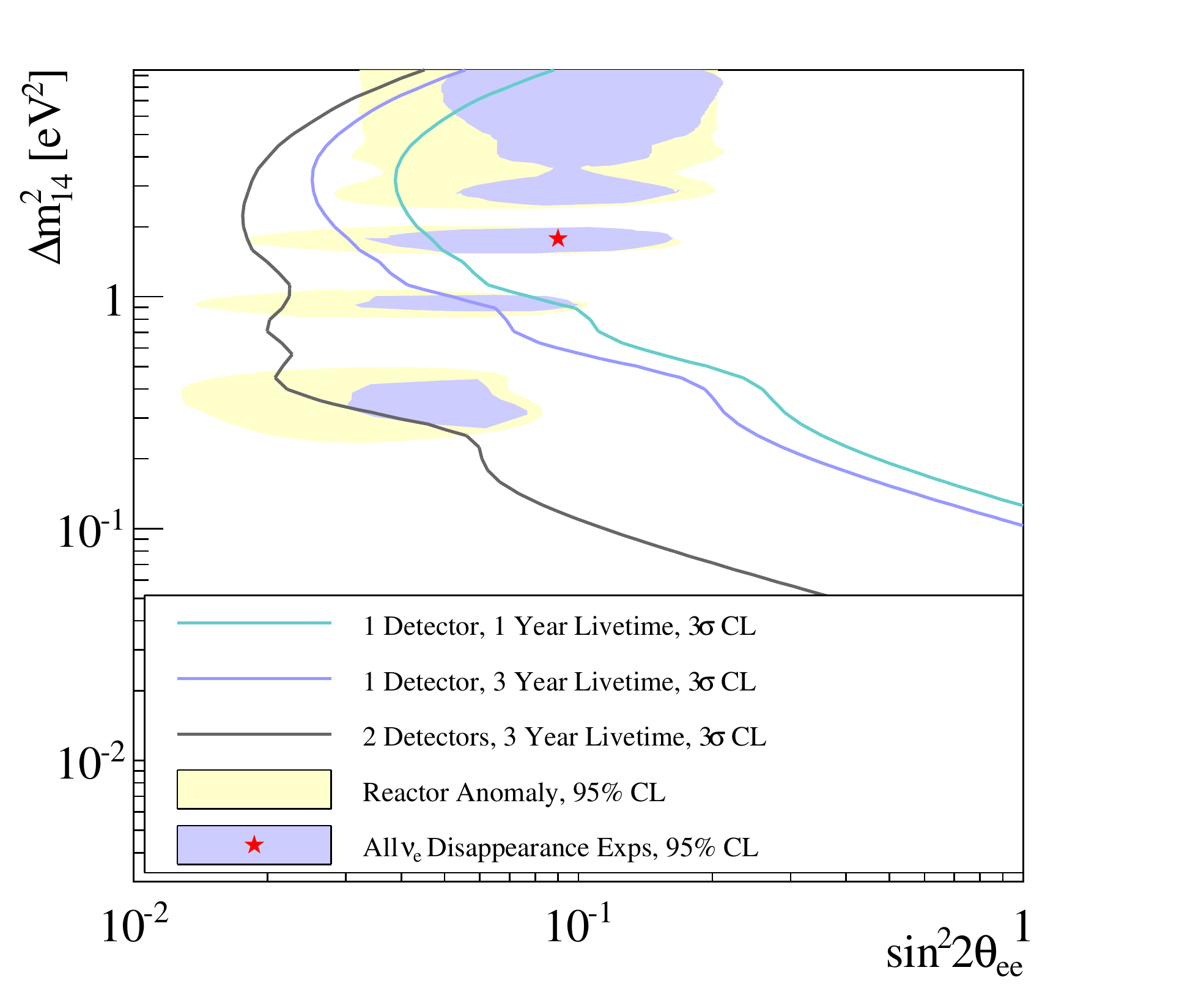}
\caption{Total sensitivity to 3+1 oscillations for Phase~I (near detector only, 1 and 3 years data taking) and Phase~II (near and far detectors, 3 years data taking).  
Excellent coverage of the phase space suggested by fits to anomalous $\nu_e$ and $\nuebar$ disappearance results and the ``reactor anomaly'' in~\cite{Kopp:2013} is achieved.  The current 3+1 best-fit (star) is addressed at high confidence level with a single year of Phase I running.}
\label{fig:chisq}
\end{figure}

By performing the $\chi^2$ fit described above, Phase I and Phase II confidence-level (CL) contours for observation of sterile neutrino oscillations have been obtained, and are displayed in Figure~\ref{fig:chisq}.  A single calendar year of data with one detector will yield a 3$\sigma$ test of the best-fit parameter space region favored in global 3+1 fits to all $\nu_e$ and $\nuebar$ data~\cite{Kopp:2013}.  High-CL exploration of portions of the suggested parameter space should be possible with less than a year of data.  If a single detector is deployed for three years, a majority of the suggested parameter space above 1~eV$^2$ can be addressed at high confidence level.  It should be stressed again that the displayed contours are totally independent of any absolute reactor flux or spectrum information, and arise from purely relative comparisons between differing locations within a single detector.  Improved reactor constraints would significantly improve these contours at all $\Delta m^2$ values.


The resultant increase in the sensitive range of $\Delta m^2$ and $\theta_{14}$ going from Phase~I to Phase~II is illustrated in Fig.~\ref{fig:chisq}.  At low $\Delta m^2$ values, sensitivity is improved as long oscillation wavelengths are resolved within the experiment's wider $L/E$ coverage.  Sensitivity is also improved at intermediate and higher values of $\Delta m^2$ as additional $L/E$ coverage allows the detection of multiple neutrino oscillation periods.

In both Phase I or Phase II, sub-dominant features in the sinusoidal $L/E$ oscillation pattern will result from the existence of multiple eV-scale neutrinos or other non-standard neutrino interactions.  A detailed demonstration of the ability to probe 3+2 sterile neutrino oscillations, and to distinguish 3+1 and 3+2 mixing with Phase~I and Phase~II is presented in~\cite{MultSBL}.

While not presented here, 3+1 sensitivity has also been investigated for the INL and NIST reactor sites. As will be discussed in Sec.~\ref{sec:USRx}, different near and far detector baselines and sizes are feasible at each site.  In addition, reactor powers and duty factors vary between sites.  Taking into account these site-to-site variations while assuming similar background conditions, it is found that roughly comparable oscillation sensitivities are achievable for Phase I and Phase II at all reactor sites.




\subsection{Direct Measurement of Antineutrinos from Spent Fuel}

During a small but significant percentage of time, the various research reactors are shut down for refueling and maintenance.  During this time, while fission has largely ceased, beta decay of fission products continues in the spent fuel in the reactor, leading to production of ``spent fuel'' antineutrinos.  A full three-year data set for either experimental phase will contain a sizable number of such events.  The measurement of the spectra and rate of these neutrinos as a function of time can provide constraints on models describing antineutrino production in reactors, and would provide the first positive measurement of remote spent fuel detection for non-proliferation purposes~\cite{HuberKorea}.  At some of the candidate sites, antineutrinos created in spent fuel repositories adjacent to the main reactor core may also be statistically accessible.  While spent nuclear fuel antineutrino statistics will be sizable for these reactor-off periods, excellent background reduction and characterization will be the key to making a statistically significant measurement with PROSPECT.

\begin{table*}[htb!pb]{%
    \begin{tabular}{c|c|c|c|c|c|c} \hline
      Site & Power (MW$_{th}$) & Duty cycle & \multicolumn{2}{c|}{Near detector} & \multicolumn{2}{c}{Far detector} \\
       & & & Baseline (m) & Avg. flux & Baseline (m) & Avg. flux\\ \hline
       NIST & 20 &68\%& 3.9 & 1.0 & 15.5 & 1.0\\ 
       HFIR & 85 &41\%& 6.7, 8.0 & 0.96, 0.68 & 18 & 1.93\\ 
       ATR & 120 &68\%& 9.5 & 1.31 & 18.5 & 4.30\\ \hline
    \end{tabular}}
  \caption{Reactor parameters and potential detector baselines for HEU research reactor facilities under consideration for PROSPECT.  The two values given for the HFIR near location are for the two available deployment locations.}
  \label{tab:rxsites}
\end{table*}

\section{Research Reactor Sites in the U.S.}
\label{sec:USRx}

The Idaho National Laboratory (INL), Oak Ridge National Laboratory (ORNL), and National Institute of Standards and Technology (NIST) operate powerful, highly compact research reactors. Each of these sites have identified potential locations for the deployment of multiple compact antineutrino detectors at distances between 4 - 25~m from the reactor cores. In this section we describe the characteristics of each of these facilities and how a short-baseline reactor spectrum and oscillation experiment can be conducted for each case.  The High Flux Isotope Reactor at ORNL has been selected as the location for Phase I of the PROSPECT experiment; however, since all three sites provide excellent characteristics for precise physics measurements and may be utilized in subsequent phases of the PROSPECT experiment, each will be described in detail below.

\begin{figure}[htb!pb]
\centering
\includegraphics*[width=0.5\textwidth]{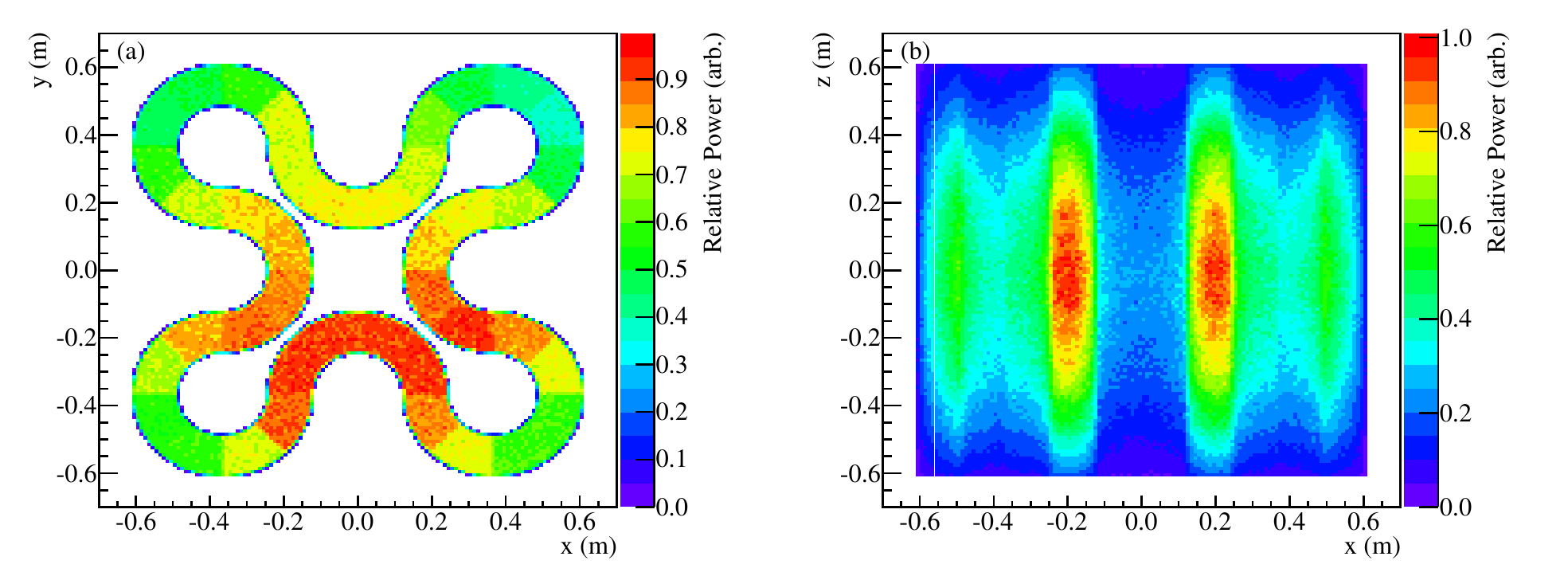}
\includegraphics*[width=0.5\textwidth]{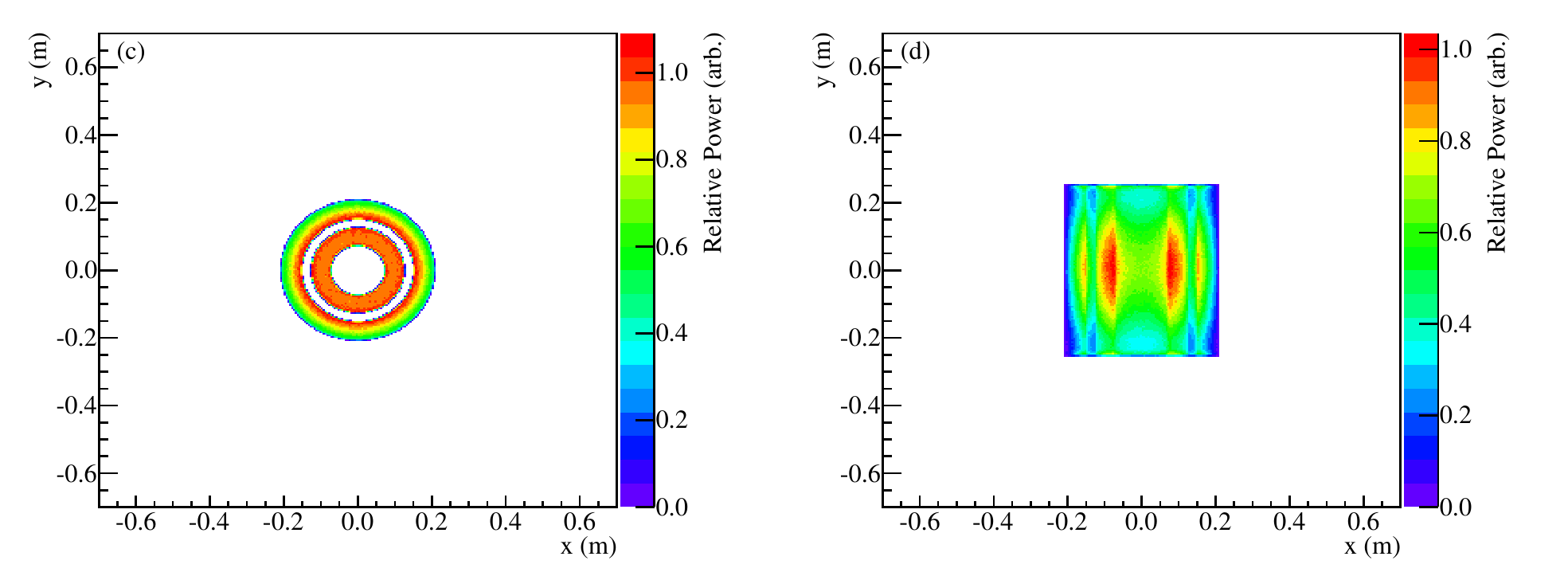}
\includegraphics*[width=0.5\textwidth]{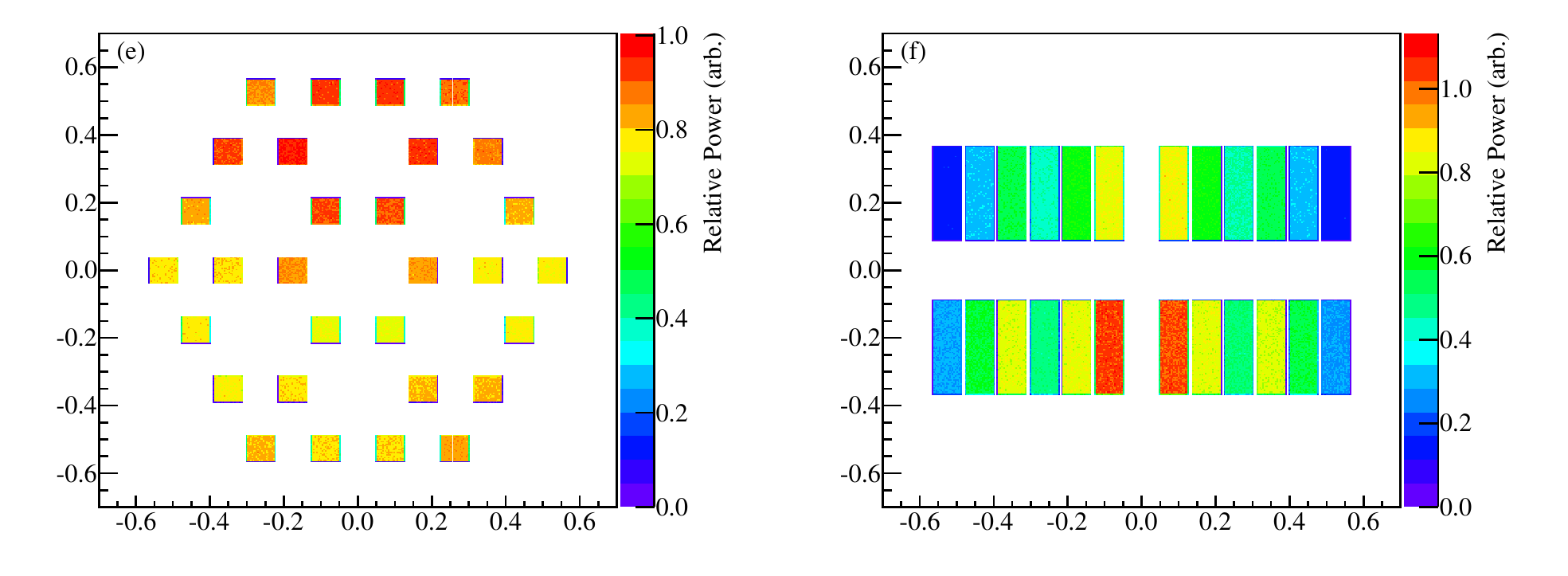}
\caption{Radial (left) and axial (right) core shapes and power distributions of U.S. research reactors: (a,b) ATR; (c,d) HFIR; (e,f) NIST. Note that the the ATR and HFIR power distributions can change slightly from cycle-to-cycle depending upon the material begin irradiated within those cores, whereas, as a dedicated neutron source, the NIST power distribution is very similar cycle-to-cycle. Each reactor site has well established evolution codes to predict and track these distributions between and within reactors cycles.}
\label{fig:rxCore}
\end{figure}

\begin{figure}[htb!pb]
\centering
\includegraphics*[width=0.45\textwidth]{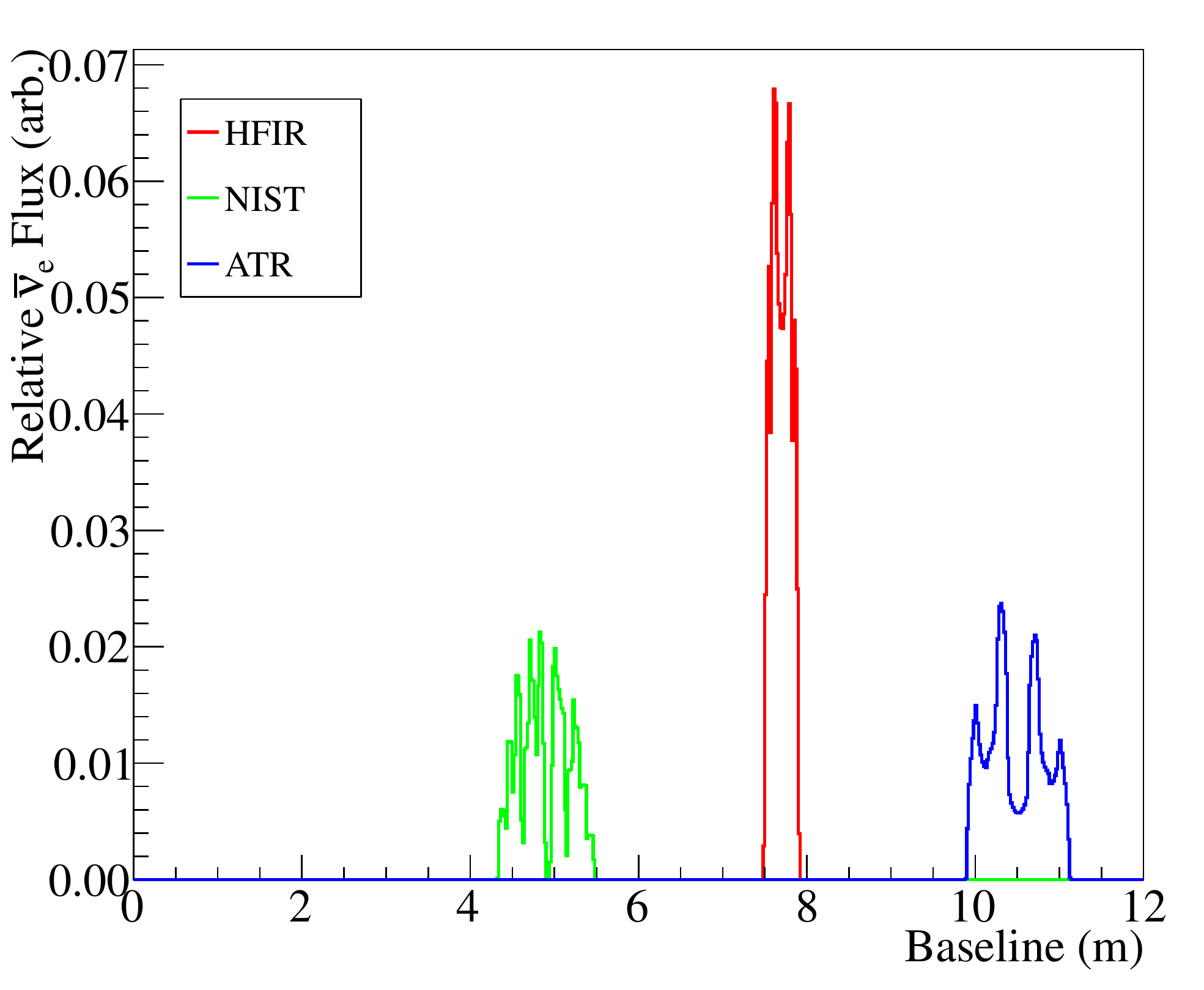}
\caption{(a) Baseline distributions for the ATR, HFIR and NIST research reactors, as viewed from the center of possible near detector locations. These distributions include the effect of solid angle and the core power distributions.}
\label{fig:rxBaselines}
\end{figure}

Reactor and site parameters relevant to PROSPECT are summarized in Table~\ref{tab:rxsites}.  The core dimensions of each of these reactors are compared in Fig.~\ref{fig:rxCore}. The diversity of shapes and sizes reflect the different functions that these facilities were designed for. The core shape combined with the physical layout of each facility determines the range of baselines that reactor-emitted $\overline{\nu}_e$ would traverse before reaching possible detector locations.  This distribution of baselines is illustrated in Fig.~\ref{fig:rxBaselines}, utilizing the reactor and site information from Table~\ref{tab:rxsites}.

\begin{figure}[htb!pb]
\centering
\includegraphics*[width=0.45\textwidth]{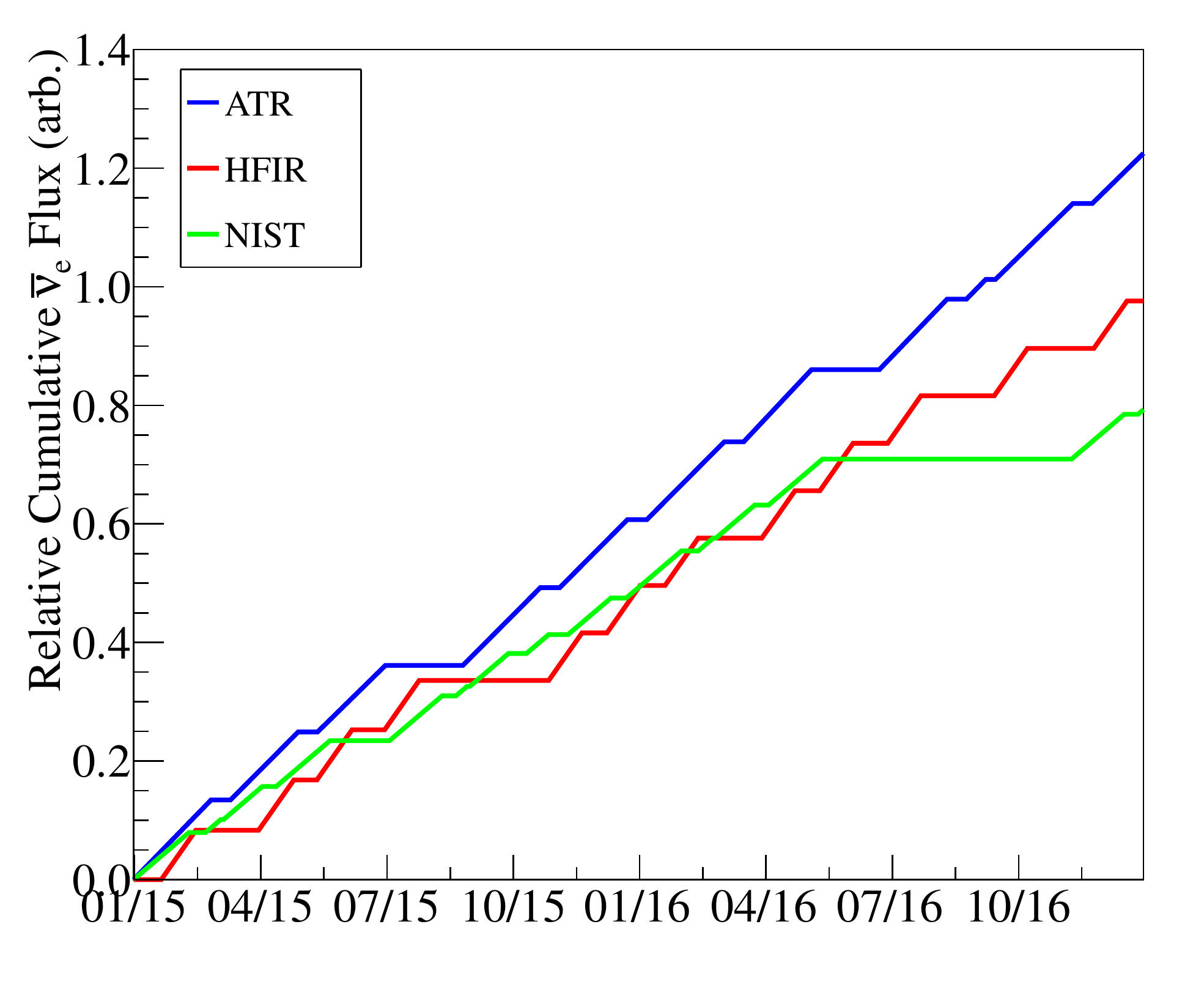}
\caption{Representative power histories for the ATR, HFIR and NIST reactors are used to generate cumulative flux curves for the near detector locations at these sites over a two year period beginning in 2015. Note that the NIST expects to conduct a maintenance outage of approximately 6 month length in 2016. The exposure obtained at each site is comparable over the expected duration of the experiment.}
\label{fig:rxPower}
\end{figure}

These facilities operate on well-planned schedules, and their central mission is to provide high reliability to many users. While the details of these operating schedules differ from facility to facility based upon maintenance and refueling needs and resource constraints, the time-averaged $\overline{\nu}_e$ flux at possible near detector locations is expected to be remarkably similar at each over the next several years (Fig.~\ref{fig:rxPower}).

\subsection{The High Flux Isotope Reactor at ORNL}

The High Flux Isotope Reactor (HFIR) was designed to provide a very high neutron flux for irradiation and isotope production applications. Multiple locations exist in the reactor for performing sample or target irradiations, some of which can be accessed during reactor operation. The HFIR core design is very compact and comprises a single aluminum fuel assembly which has two annular fuel elements consisting of thin inserts of U$_3$O$_8$ (Fig.~\ref{fig:rxCore}c).  The full assembly is replaced after each operating cycle.

HFIR operates at a consistent power of  $85$\,MW$_{th}$, with occasional operation at lower powers during the short ($\sim$hours) cycle startup phase. A reactivity control system maintains this power throughout the cycle, irrespective of the irradiation experiments being performed. Given this consistent operation, it is not typical to perform as-run reactor simulation analyses cycle by cycle, although a detailed Monte Carlo N-Particle (MCNP) model of the core is available for this purpose.

HFIR cycles last approximately 25\,days with deviations from that average being less than 36\,hours. At present, 6\,cycles per year are scheduled, giving a duty cycle of $\sim41\%$.  Outages between cycles  have at least 14\,day duration, with generally one longer outage per year.  No extended outages beyond those described above are currently planned.  In compliance with global non-proliferation treaties, HFIR is scheduled to convert to LEU fuel, but this will be phased over several years and will not commence until at least 2020.

A potential near detector deployment location has been identified $\sim4$\,m above the reactor centerline along a hallway bordering the reactor pool shielding structure, with access directly provided by large exterior doors at the end of the hallway.  The region between the core to the closest wall in the confirmed near detector location consists of light water in the reactor pool surrounding the core and thick concrete pool walls, with a total core center to wall exterior distance of~5.5~m.   At this location, a 3.0~m width x 2.7~m height x 3.3~m length shielding/detector package is deployable at baselines of $\sim$5.7-11.1~m.  Fire egress requirements constrain detector and shielding deployments to either the extreme front or back of this range, allowing for two possible near detector deployment locations, termed the 'front' and 'back' near locations.  Overburden is provided by the structure of the reactor building, consisting of 0.5~m or less of concrete. A potential far detector location is located exterior to the reactor building approximately in-line with the core center that provides a range of baselines between 16-24\,m.

The proposed near detector deployment location has already been utilized by the collaboration for a variety of purposes, including the deployment of particle detectors for background surveys, which will be described in Section~\ref{sec:bkg}.  These locations are restricted but fully accessible to properly trained collaborators during both reactor-on and reactor-off periods.  HFIR data networks are restricted but accessible to collaborators and have been utilized for remote experiment run configuration and data transfer.

The HFIR reactor site has been selected for Phase I of the PROSPECT experiment.  The confirmed HFIR detector location allows the deployment of a comparatively large detector/shielding package at a short and broad range of baselines in a vicinity largely free of highly-time-dependent reactor-related backgrounds, making it an excellent environment for a precise oscillation search and spectral measurement.  In addition, HFIR is a user facility that is accustomed to hosting visitors and scientists within their complex.  Ease of access to the lab and to the confirmed detector location, access to existing data networks, and significant on-site participation of engineering and science staff provide an environment conducive to efficiently carrying out needed R\&D and experimental tasks.

\subsection{The Advanced Test Reactor at INL}

The Advanced Test Reactor (ATR) was designed to support a wide variety of materials and system investigations. The ATR design exploits a unique serpentine core configuration to offer a large number of in-core positions for testing (Fig.~\ref{fig:rxCore}a). The core is comprised of $40$\,HEU fuel assemblies, approximately one third of which are replaced after each cycle. The typical residency of an assembly in the core is 2-3 operating cycles. The operating power of  ATR is in the range $110-120$\,MW$_{th}$, although occasionally short cycles operate as high as $200$\,MW$_{th}$.

The operating power and core power distribution vary from cycle-to-cycle. The unique design of ATR permits large power variations among its nine flux traps using a combination of control cylinders (drums) and neck shim rods. Within bounds, the power level in each corner lobe of the reactor can be controlled independently during the same operating cycle. Following each cycle, Òas-runÓ analyses based on in-core measurements and reactor simulations can provide more precise power estimates for each area of the reactor.

ATR typically operates on a schedule with approximately 50-60\,days at power then 15-30\,days with the reactor off. There are a few exceptions to this schedule. Approximately every 2 years there is a 3-4\,month outage and every 10\,years a 6-8\,month major outage in which internal core elements are replaced. The next such replacement outage is proposed for Apr.-Oct.\,2017. ATR is also scheduled to convert to LEU fuel after 2020.

The top of the ATR reactor vessel is approximately at grade, while the center of the reactor core is located approximately $5.5$\,m below grade. Sub-basement levels of the facility contain potential antineutrino detector deployment locations, with access being provided by a large overhead crane, a freight elevator and wide corridors. Potential near detector locations have been identified in the first sub-basement of the ATR building. The floor of this level is $5.8$\,m below grade, placing it approximately in line with the core center. At least $3.3$\,m of concrete and $1$\,m of water lie between the reactor core and this location. The distance from the core center to the closest wall in this location could be as little as $7.9$\,m, while the center-to-center distance from the core to the nominal detector configuration discussed below is $9.6$\,m. While the location is below grade, the overburden is primarily provided by building structure including a  concrete floor of $\sim20$\,cm thickness and the exterior structure of the ATR building.

A potential far detector location has been identified in the second sub-basement of the ATR building. The floor of this level is $11.6$\,m below grade and at least $5.5$\,m of concrete and $1$\,m of water lie between the reactor core and this location. The center-to-center distance from the core to the nominal far detector configuration discussed below is $18.5$\,m. In this location overburden is primarily provided by building structure including two concrete floors of $\sim40$\,cm total thickness and the exterior structure of the ATR building.

\subsection{The National Bureau of Standards Reactor at NIST}

The National Bureau of Standards Reactor (NBSR) at NIST is a heavy water (D$_2$O) cooled, moderated, and reflected, tank-type reactor that operates at a designed thermal power of 20 MW.  The NBSR is fueled with high-enriched U$_3$O$_8$ in an aluminum dispersion that is clad in aluminum.  The HBSR core comprises 30 HEU fuel assemblies arrayed in an approximately cylindrical geometry (Fig.~\ref{fig:rxCore}e).  As with the other sites, NIST is scheduled to make a phased conversion to LEU fuel sometime after 2020

A reactor cycle is nominally 38 days.  The startup usually can be accomplished in about 2 hours, after which the power is  maintained at 20 MW for the remainder of the cycle. Variations in power are minimized by the automatic movement of a regulating rod.  A detailed MCNP model of the core has been used to model power distribution as a function of time and this data is publicly available.  Aside from the normal operating schedule there is a longer six-month shutdown planned for mid-2016.

Multiple potential detector locations have been identified within the NBSR confinement building as well as two sites outside and adjacent to the confinement building.  The shortest available baseline, allowing for an occupied space of approximately 2.0~m  wide by 3.0~m high by 3.5~m long, including shielding, is just outside a segment of the biological shielding at an instrument station designated the ``thermal column''.  The thermal column consists of a heavy water tank (now filled with light water) and graphite block shielding that was previously used as a facility for intense thermal neutron beams.  Due to the shielding design, fast neutron backgrounds are expected to be lower in this region than elsewhere in the confinement building.  The face of the shielding to the center of the core is approximately 3.5 m.  The regions on either side of the identified deployment location are occupied by neutron scattering experiments which produce ambient neutron fluxes that vary continuously based on the operational status and run mode of each machine.
Two far detector locations have been identified, both with baselines of roughly 16 m.  All locations are at grade and roughly in-plane with the core.  Far locations have no overburden, while near locations are under roughly 50 cm of concrete that comprises the structure of the confinement building.

The location of the near detector is  accessible though a loading dock.  The area is serviced by a 15 ton radial crane.  
Limited deployment at the thermal column site for prototyping purposes is possible immediately.  

\section{Expected Backgrounds at Near-surface Research Reactor Sites}
\label{sec:bkg}
The sites detailed in Section~\ref{sec:USRx} have relatively little overburden and will require the operation of detectors close to the reactor core where both fast-neutron, neutron-capture and gamma-ray backgrounds are potentially high. Therefore, both fast neutron and muon fluxes through an unshielded detector are expected to be several orders of magnitude higher than the neutrino detection rate. 

In an IBD analysis, background events can manifest in two important ways:
\begin{itemize}
\item \textit{Neutron-capture Correlated Backgrounds}
Events involving one or more neutrons, resulting in an inter-event time correlation similar to that observed for IBD.   Fast neutrons, for example, can emulate an IBD event through proton scatter and subsequent recoil followed by thermalization and delayed  capture.  Similarly, multiple neutrons resulting from the same cosmogenic particle may capture at different times, mimicking an IBD event. 
\item \textit{Random Coincidence Uncorrelated Backgrounds}
Random coincidences involving primarily either neutron recoils, neutron captures, or gamma-ray interactions can mimic an IBD event should they occur  with an appropriate energy range and within the average neutron capture time.
\end{itemize}

\begin{figure*}[tb]
\centering
\includegraphics*[width=0.75\textwidth]{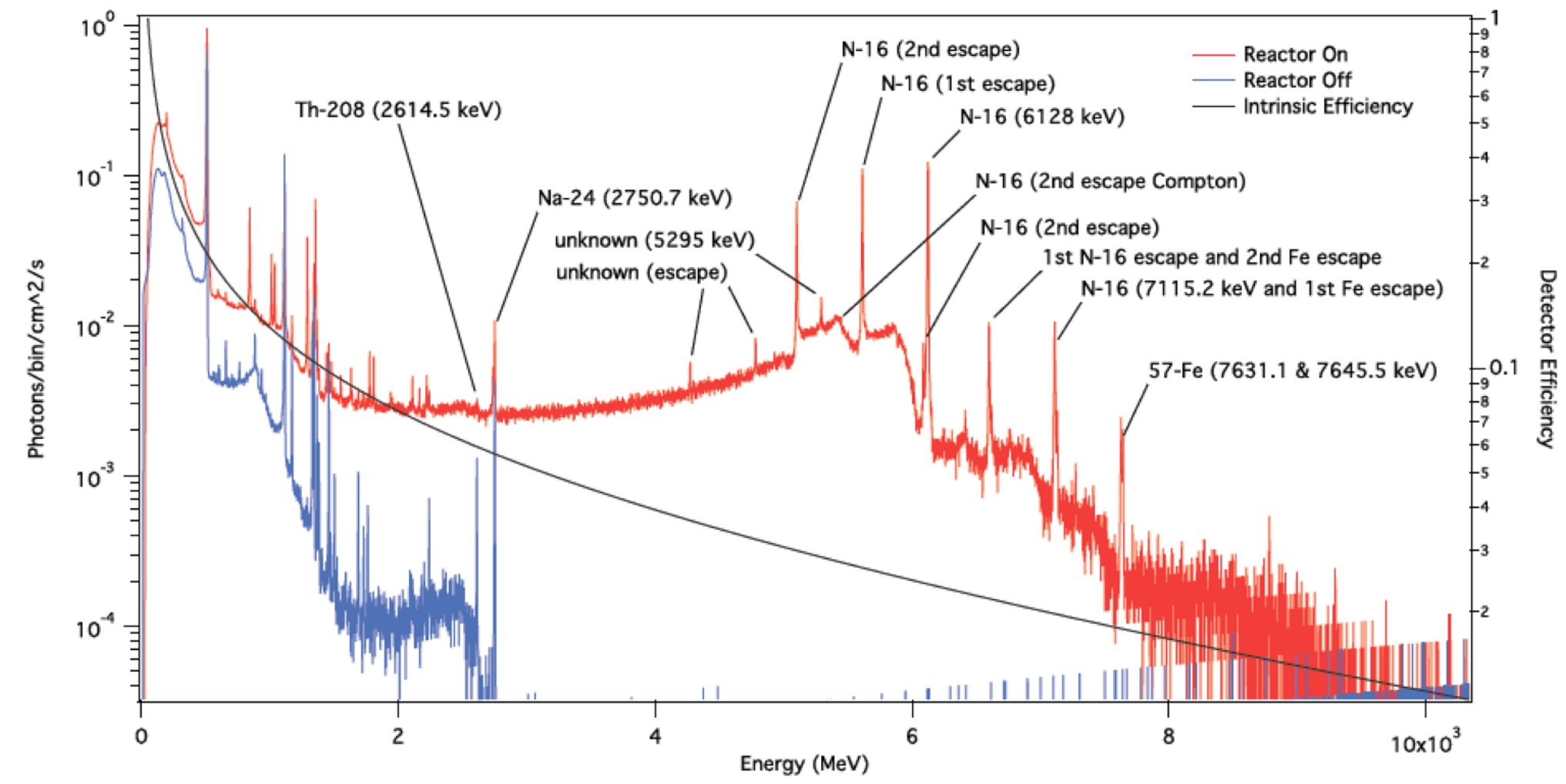}
\caption{Germanium spectrum taken at the potential near location at NIST during reactor on and off periods. The reactor on spectrum represents the current shielding configuration of adjacent experiments.}
\label{fig:Gammabkg}
\end{figure*}

The backgrounds that will be encountered at a research reactor site fall into several broad categories, based upon their source and they way in which they manifest in a $\overline{\nu}_e$ detector.  These backgrounds will be generally described below, as will completed and current efforts to survey the existing background environment in the vicinity of the identified PROSPECT deployment locations at each site.  An overview of completed survey measurements is given in Table~\ref{tab:BkgSurveys}.  A detailed quantitative summary of background surveys at the various sites is currently being organized for publication.  Background measurements currently underway at HFIR utilizing PROSPECT prototype detectors will be be overviewed in Section~\ref{sec:rd}.

\begin{table}[!htb]{%
    \begin{tabular}{l | l | l | l | l} \hline
      \multicolumn{2}{c|}{Survey Type} & HFIR & ATR & NIST \\ \hline
      \multirow{4}{*}{Gamma} & High-Res (LaBr, HPGe) & X & X & X \\
      & Medium-Res (NaI) & X & X & X \\
      & Directional/Spatial Variation & X & X & X \\
      & Adding Localized Shielding & X &  &  \\ \hline
      \multirow{3}{*}{Neutron} & Bonner Ball & X & X & X \\
      & FaNS Detector & X &  & X \\
      & Stilbene Detector & X & X & X \\ \hline
      \multirow{1}{*}{Muon} & Muon Detector & X & X & X \\ \hline
    \end{tabular}}
  \caption{Completed background survey measurements for the considered three reactor sites. 'X' indicates a completed measurement.  Quantitative summaries of acquired data will be given in a subsequent publication.  Other measurements of backgrounds with prototype PROSEPCT detectors will be discussed in Section~\ref{sec:rd}.}
  \label{tab:BkgSurveys}
\end{table}

\subsection{Reactor Correlated Backgrounds}
Locating a detector in close proximity to a reactor core is likely to introduce the special challenge of reactor correlated backgrounds that are likely to vary in time as well as space.  Both high-energy gammas and fast neutrons can contribute significantly to backgrounds in detecting inverse beta decay.

\subsubsection{Fast Neutrons}
Fast neutrons are copiously produced by fissioning isotopes and emitted by an operating nuclear reactor.  Despite significant neutron shielding around a core, some fraction of produced fast neutrons will be present in proposed deployment locations.  At some locations, $e.g.$, NIST, fast neutron backgrounds are dominated by partially thermalized fission neutrons scattered from adjacent experiments.  A suite of comparative fast neutron measurements between reactor sites and locations, listed in Table~\ref{tab:BkgSurveys} and described below, have been conducted with various neutron detectors.

Bonner balls are a commonly used in medical physics to determine radiation dosage related to neutron exposure.  Measurements with calibrated Bonner balls at HFIR yield a fast neutron flux of 2-3 cm$^{-2}$\,s$^{-1}$ likely peaked in the 1-2 MeV range.  Deployments of identical Bonner ball setups at all three sites for comparison of neutron fluxes are also complete and exhibit results of similar magnitude.

Two segmented Fast Neutron Spectrometers (FaNS) have been developed at NIST and the University of Maryland, and have been used to further characterize fast neutron backgrounds~\cite{Langford:2014upa, tomthesis:2013}. The FaNS detectors are capture-gated spectroscopy arrays of plastic scintillator and $^3$He proportional counters. 
These detectors have been calibrated in mono-energetic neutron fields and have also been used to measure the cosmogenic neutron spectrum at the surface. Both show good agreement with MCNP predictions, including the surface spectrum up to 150~MeV. Deployment of FaNS and subsequent data-taking has been completed in various locations at both ORNL and NIST, including near and far detector locations as well as reactor-on and reactor-off periods.  

Additionally, a small neutron detector containing  a 2'' trans-stilbene crystal developed at Lawrence Livermore National Laboratory was also deployed at all three reactor sites.   Through pulse shape discrimination, this detector is capable of distinguishing nuclear recoils from electromagnetic interactions, allowing a precise relative comparison of fast neutron fluxes between all three sites even in the presence of significant high-energy gamma backgrounds.

\subsubsection{High-Energy Gamma Radiation}

An example of the gamma-ray background that can be encountered at a research reactor is shown in Fig.~\ref{fig:Gammabkg}. Here, a High Purity Germanium (HPGe) gamma-ray spectrometer has been used to conduct a detailed survey of the potential NIST near detector deployment location. Backgrounds above 2.4~MeV are dominated by thermal neutron capture on reactor and experiment structural and cooling materials, yielding prompt gammas from $^{16}$N and $^{57}$Fe at 6.1~MeV and 7.6~MeV respectively. The reactor correlated component of the gamma-ray background is evident from the comparison of reactor-off and reactor-on spectra shown in Fig.~\ref{fig:Gammabkg}.  Notably, in completed background surveys at NIST, the $^{16}$N flux shows a clear angular dependence consistent with it originating in water filled pipes visible from certain portions of the near detector location.  

These general observations highlight the need for targeted spatial, directional, and temporal surveys of gamma backgrounds at intended detector deployment locations.  Surveys have been done with high-resolution HPGe or LaBr spectrometers (NIST and ATR respectively), and with the same moderate-resolution NaI spectrometer at all three sites.  At HFIR and to a lesser extent at NIST and ATR, NaI directional gamma measurements have been done at a wide variety of positions at the near detector deployment location to identify dominant localized sources of gamma backgrounds.  Subsequent additional gamma surveys were then completed for a variety of localized gamma shielding configurations.

\subsection{Natural Radioactivity Backgrounds}
In most $\nuebar$ detectors, gamma, beta, and alpha decay products of the U, Th, and K decay chains present in doped scintillator, photomultiplier (PMT) glass, and metal building materials surrounding the active detector target can interact within the target scintillator, producing mainly isolated, low-energy triggers.  These plentiful triggers can randomly overlap in time with uncorrelated neutron interactions with the target, giving a signal-like time-coincident signature.

These radioactive background triggers can be reduced using now-standard precautions in neutrino physics, such as providing a non-scintillating buffer between PMTs and the detector target, purifying scintillator of radioactive contaminants during scintillator production, and radioassay of all detector components prior to detector construction.

\subsection{Cosmogenic Backgrounds}
Muon rates at all three reactor sites are high with respect to a typical underground neutrino detector, with fluxes through the active region of the detector and the shield expected of magnitude $\sim$200 Hz and 1 kHz respectively.  The hadronic component of the cosmic ray flux will also impinge the detector and shielding. In addition, both spallation and muon-capture can yield very high-energy secondary neutrons originating within the passive shielding.  Through thermalization or inelastic collisions and subsequent capture, primary or secondary neutrons can mimic IBD events.  In addition, multiple neutrons generated by the same initial cosmogenic particle can capture at different times, resulting in the same timing profile as an IBD event.  In the absence of a muon veto system, through-going muons producing spallation neutrons can also mimic the time and energy profile of IBD interactions along detector edge boundaries.  Detailed Monte Carlo simulations of these potential IBD backgrounds are in progress.

Comparative cosmic background surveys have been completed at all three sites.  A directional muon detector consisting of a stacked array of scintillator panels has been deployed to measure relative muon rates and possible shielding effects of nearby objects.  In addition, the FaNS neutron spectrometers described above provide the energy profile of neutrons up to 200~MeV, providing a relative comparison of cosmic and secondary neutrons between sites.

Spallation-produced radioisotopes such as $^8$He and $^9$Li have relatively long half-lives, 119~ms and 178~ms respectively, have Q-values of roughly 10~MeV, and beta-decay to neutron-unstable daughters.   The decay of these isotopes can thus closely mimic the IBD of a reactor antineutrino.  Rough estimates of isotope production rates are conservatively less than 1000~d$^{-1}$, indicating a challenging but tractable background.  Further detailed studies are in progress.  

\section{Detector Concepts}
\label{sec:detector}

\begin{figure*}[tb]
\centering
\includegraphics*[width=0.9\textwidth]{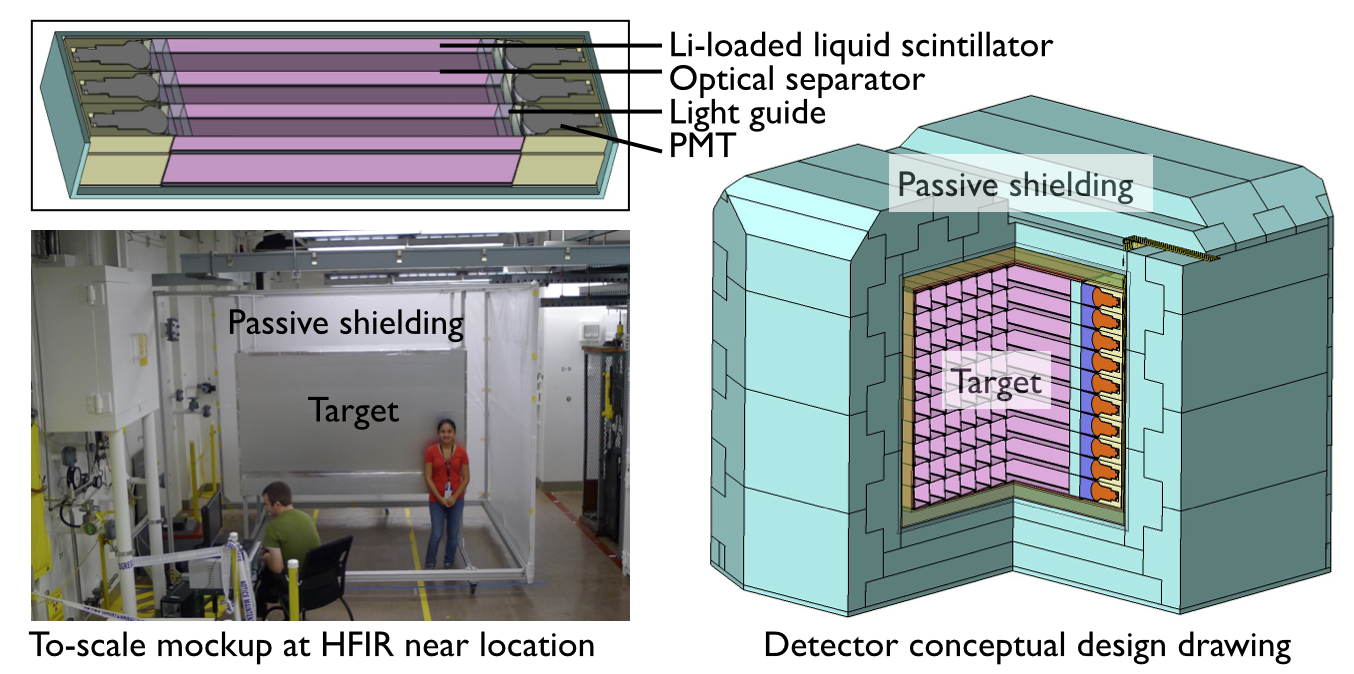}
\caption{Bottom left: a mock-up of the placement of the near detector at its intended HFIR deployment location. Right: the conceptual near detector configuration showing a segmented liquid scintillator antineutrino target and passive shielding.  Top left: a conceptual design of a group of target segments, comprised of optically separated Li-loaded scintillator volumes readout by a PMT at each end.}
\label{fig:detector}
\end{figure*}


For PROSPECT to perform a precision oscillation and spectrum measurement within $10$\,m of a research reactor its detector design will have to provide excellent background rejection  and support precise calibrations of energy, position, and relative efficiency across the active volume.  This section highlights the key features of the PROSPECT detector design essential to reaching these background-reduction and detector response goals.

\subsection{Detector Segmentation}
While recent reactor antineutrino experiments have used cylindrical homogeneous liquid scintillator designs, the PROSPECT detectors will utilize a segmented design (Fig.~\ref{fig:detector}). The basic PROSPECT target segment comprises a long rectangular optically isolated scintillator volume read out on its long ends by PMTs. There are several reasons to pursue this approach. First, the segmentation provides intrinsic position resolution sufficient for target fiducialization or an oscillation measurement in two axes. Second, the segmented approach is space-efficient, requiring optical readout on only one or two sides, an important consideration in the compact spaces available at HFIR and other reactor sites. Third, a segmented design of this type provides the opportunity to control and optimize the optical collection properties of the detector through the choice of aspect ratio and reflector material, which is potentially important for use of the pulse shape discrimination (PSD) background rejection technique. Finally, segmentation provides the ability to discriminate single- versus multi-site event topologies, e.g. highly-localized energy depositions from Li-captured neutrons, and to identify triggers coincident in both time and space, which can be used for signal identification and background rejection.

\subsection{Low-Mass Optical Separators and In-Target Calibration Source Deployment}

To maintain a high antineutrino energy resolution, a segmented target must exhibit high light collection efficiency that is relatively uniform throughout the target volume, and must absorb as little IBD positron and annihilation gamma energy as possible.  To achieve these goals, the detector target will consist of a single large tank of scintillator separated into long rectangular cells by low-mass optical separators.  As pictured in Figure~\ref{fig:detector}, rectangular cells allow for a high packing density as well as uniform light collection across the segment cross-section.  Optical separators composed of thin highly reflective films and thin, rigid scintillator-compatible support and encapsulation materials can ensure high light collection efficiency along the entirety of the cell while maintaining PSD capabilities with an in-target non-scintillating volume of a few percent or less.  Degradation of energy resolution from leakage of antineutrino-related energy can also be reduced in this design by utilizing the existing segmentation to fiducialize the target volume.

To precisely demonstrate understanding of detector energy response, calibration sources must be deployed at various locations in the detector target.  The optical separator system also accommodates this deployment by including hollow rods running the length of each segment in each cell corner edge.  In addition to providing further mechanical support for the separator structure, support rods provide a path through which calibration sources and their deployment systems can be integrated into the detector interior.  Designs for compact optical and radioactive calibration sources are currently being developed.



\subsection{Double-Ended Photomultiplier Readout}

After being produced by particle interactions in the target and propagated down the cell by high-reflectivity optical separators, scintillation light will be collected by one photomultiplier tube at each end of each PROSPECT sub-cell.  By directly collecting light at both ends of each target sub-cell as opposed to a single end, the average value and spread of photon pathlengths in the cell are reduced, which will improve both light collection and PSD performance.  In addition, double-ended readout allows for position reconstruction along a cell through relative timing and charge comparisons between PMTs.  Along with being useful for further background reduction and target fiducialization, this position information can be used to calibrate and minimize any residual differences in light collection and PSD.  

\subsection{Scintillating Detector Target With Lithium Dopant}

Scintillator loaded with a neutron capture agent is the target material of choice for antineutrino detection as it enhances the time-coincidence signature of the positron annihilation and neutron capture resulting from the Inverse Beta Decay (IBD) interaction. The scintillator dopant increases the neutron capture cross-section, shortens the capture time, and provides a more distinct signal than the single $2.2$\,MeV gamma ray emitted after neutron capture on hydrogen. Both Gd and $^6$Li doped scintillators have been used in past reactor antineutrino experiments. Neutron capture on Gd provides a distinct $8$\,MeV gamma ray signal above most natural backgrounds, when all of the gamma-ray energy released by the excitation cascade can be captured in the detector volume. However, the leakage of gamma rays near the detector edge can lead to detection efficiency variation and related systematic effects, especially in compact devices like those that will be necessary for operation near a research reactor core. By contrast, the $^3$H and $^4$He produced in the $^6$Li neutron capture reaction have a very short range, resulting in a larger and more uniform detection efficiency in a compact device when combined with pulse-shape discriminating liquid scintillator.

\subsection{Pulse-Shape Discriminating Liquid Scintillator}

The dense energy depositions produced by interaction of hadrons in liquid scintillator produce longer-lived excitations of that scintillator, leading to an elongated scintillation time profile.  When combined with with modestly fast (ns-scale) data acquisition systems, this pulse-shape discrimination (PSD) technique allows for discrimination between electromagnetic (gamma, e$^+$, e$^-$) and nuclear (p, $^4$He, $^3$H) interactions in the scintillator.  Using this technique, the heavy ion products of $^6$Li neutron capture can be identified on an event-by-event basis, providing an unambiguous indication of neutron capture. This can provide  stronger uncorrelated gamma-ray rejection than Gd-doped scintillator, as well as rejection of an important multiple neutron time correlated background produced by cosmic-ray muons.  PSD can also be used to identify time-correlated proton recoil energy depositions caused by fast neutrons. The particle identification capabilities provided by this technique are likely to play an important role in providing the background rejection required for PROSPECT.


\section{Research and Development Efforts}
\label{sec:rd}
In order to efficiently address the varied challenges discussed in the previous sections, we are developing a significant effort to demonstrate the viability of the experimental design and provide the required enabling technologies. Accordingly, the present focus is on demonstrating sufficient background rejection through the development of optimized shielding, optimized detector segmentation, and Li-doped liquid scintillators with high light yield and good pulse shape discrimination.  Significant effort is also underway to control and understand variations in detector segment response.  A key part of this process is the assembly, testing, and deployment of a suite of prototype PROSPECT detectors at the HFIR near detector location.  Here we discuss the most important aspects of the completed and current R\&D and prototyping efforts.

To meet PROSPECT's physics goals, a lithium-loaded scintillator that has excellent stability and light yield, provides efficient antineutrino detection and background rejection, and has a high flash point is required.  Acceptable light yield and PSD capability have been developed through examination of differing scintillator base, fluor, and wavelength-shifter combinations.   We have also developed compounds and techniques to support $^6$Li loading in high flashpoint solvents, some of which are described in~\cite{Bass:2012ur, Zaitseva2013747}.  

\begin{figure*}[htb!pb]
\centering
\includegraphics*[trim=0.1cm 0.1cm 1.2cm 0.1cm, clip=true, width=0.55\textwidth]{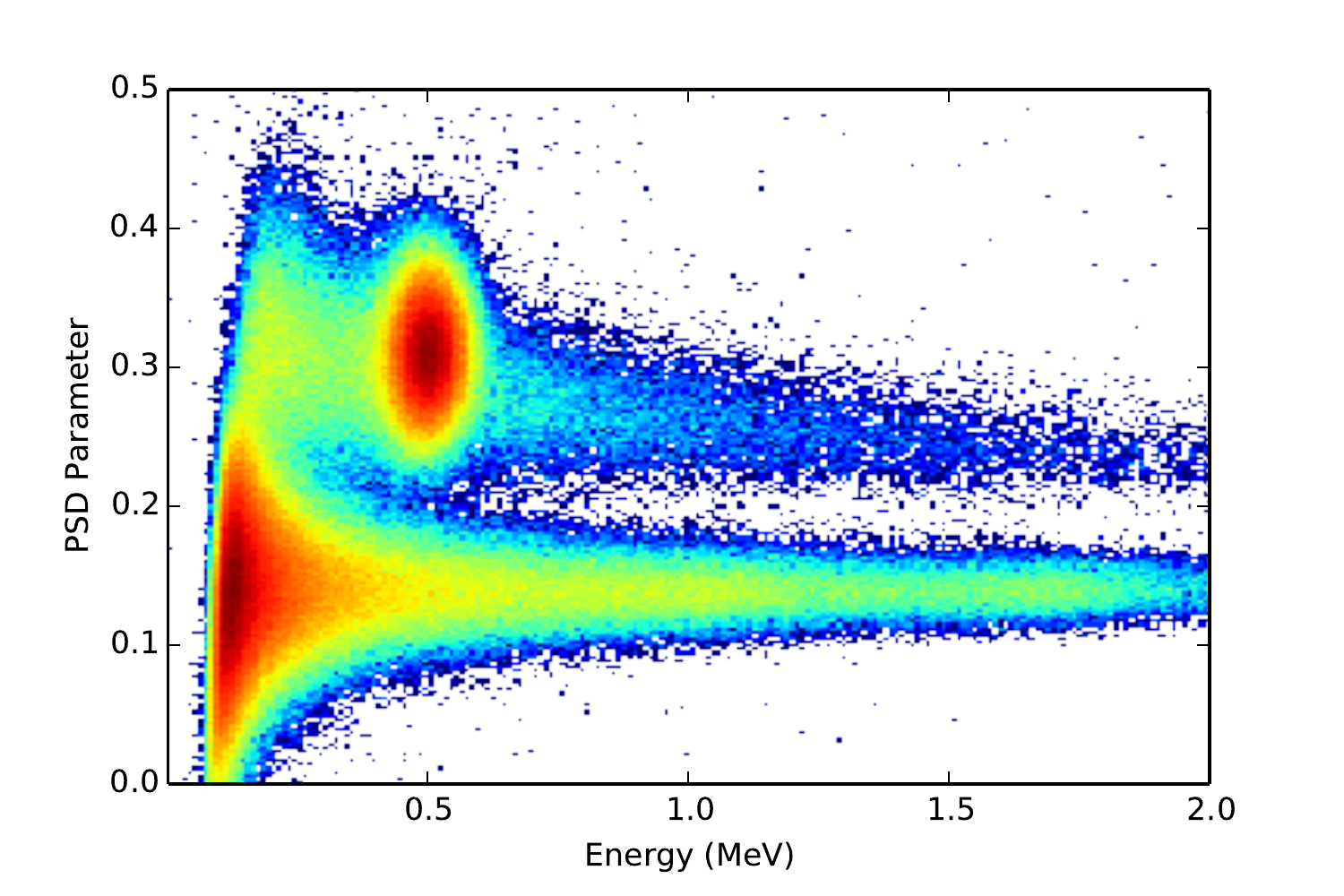}
\includegraphics*[trim=0.1cm 0.1cm 1.2cm 0.1cm, clip=true, width=0.44\textwidth]{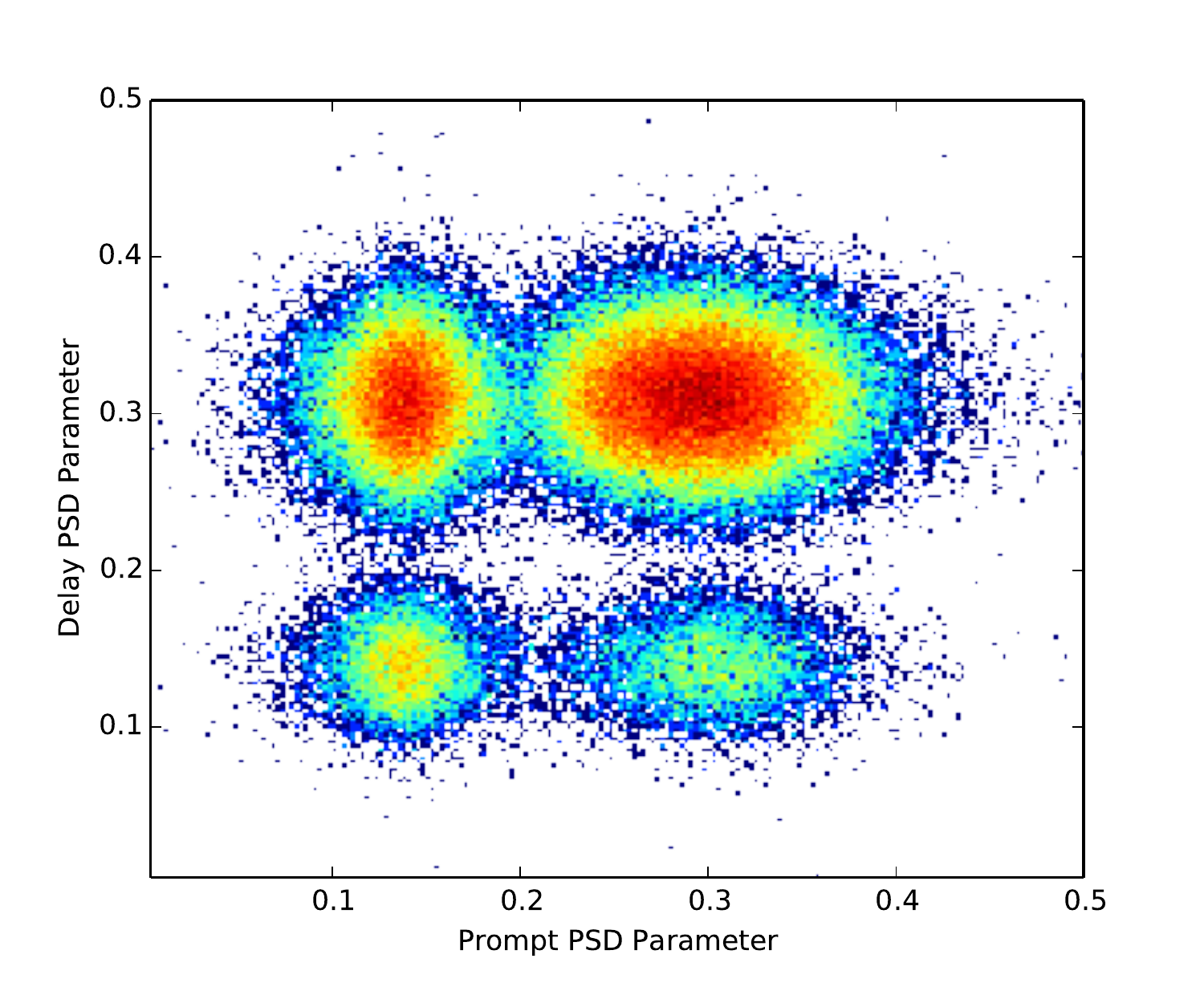}
\caption{Left: PSD parameters and energies of measured $^{252}$Cf spontaneous fission gammas and neutrons in a PROSPECT LiLS test cell.  The Li capture peak is prominent at low energies and high PSD, as is the significant gap in PSD between gamma (bottom band) and neutron-related (top band) energy depositions.  Left: PSD parameter values for prompt and delayed signals in detected time-coincident triggers in the same test cell.  Separation between IBD-like coincidences (top left), accidental gamma coincidences (bottom left), and fast- or multiple-neutron related coincidences (top right) is clearly visible.}
\label{fig:psd}
\end{figure*}

The collaboration has developed three differing cocktails that meet the general criteria described above.  Figure~\ref{fig:psd} provides a demonstration of the achieved pulse-shape discrimination ability for a small cell containing one of the developed Li-loaded scintillators.  One can clearly identify the Li-capture peak produced when exposed to a $^{252}$Cf neutron source, as well as the significant discrimination power between this peak and other gamma-related backgrounds in the cell.  The separated neutron-like and electron-like bands are also clearly visible, which will allow for significant rejection of multi-neutronand fast-neutron backgrounds.  Further details on these scintillators will be provided in an upcoming publication.  Work continues in characterizing the properties of these scintillators, particularly their long-term stability and their compatibility with other designed target materials.

\begin{figure}[htb!pb]
\centering
\includegraphics[trim=0.1cm 0.1cm 1.2cm 0.1cm, clip=true, width=0.48\textwidth]{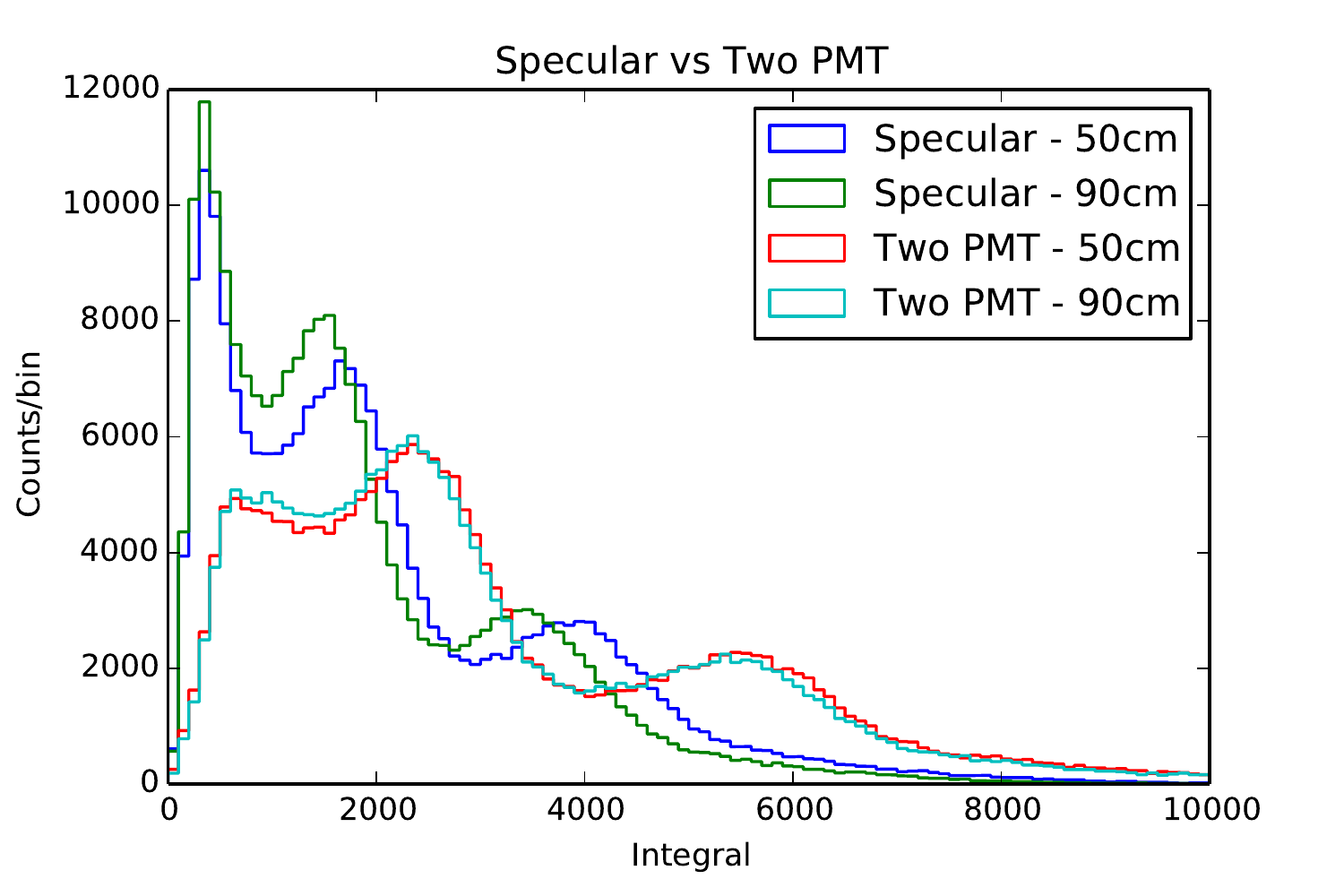}
\caption{Waveform integrals for triggers from a gamma calibration source deployed along a 1~m long PROSPECT scintillator test cell.  Data is shown at 50~cm (cell middle) and 90~cm (cell end) along the cell length for either double-ended ('Two PMT') readout or single ended ('Specular') readout, with all non-read-out ends covered in specularly reflecting film.  Light collection is higher and more stable with variation in source deployment location for double-ended readout.}
\label{fig:DoubleEnded}
\end{figure}

To develop a detector with acceptable PSD performance and light collection, appropriate segment geometry, reflector and PMT selection, and DAQ design are also of crucial importance (see, e.g., \cite{Matsuyama1996246,Matsuyama1997439}).  The detector response and background rejection capabilities accompanying various detector segmentation and readout styles are being studied via simulation and benchtop measurement.  To give one example, Figure~\ref{fig:DoubleEnded} shows benchtop data investigating differences in response between single- and double-ended cell readout.  For this data, a radioactive calibration source is deployed externally at the center and end of long rectangular cell with either single- or double-ended PMT readout and specularly reflecting walls.  In the case of one-ended readout, light collection varies by roughly 10\% as the source is moved from cell center to end.  For two-ended readout, overall light collection is significantly higher and nearly identical for the two different source positions.  With additional position reconstruction corrections in the double-ended case, light collection variation and attendant energy resolution contributions can be reduced to the percent level.  
Similar benchtop studies investigating light collection and PSD variations with these and other parameters, such as wall reflection and PMT selection, will be discussed in detail in an upcoming publication.  Complimentary simulation studies utilizing single- or multiple-cell configurations provide Monte Carlo-based estimates of these same properties as well as topology-based background reduction capabilities.

Well-optimized passive shielding will be needed in addition to the active background rejection techniques discussed above. In practice, the amount of shielding that can be used is limited primarily by constraints on space and weight. Fast neutrons below a few MeV and thermal neutrons can be sufficiently suppressed with careful design. For example, a factor of $10^{-6}$ suppression of neutrons at $1$\,MeV can be attained with $\sim$0.6~m of polyethylene, while hermetic boron-loaded shielding efficiently eliminates thermal neutrons.  Attenuation of prompt neutron-capture gamma-rays requires high-Z materials such as lead.  Optimization of such shielding based on GEANT and MCNP models, subject to realistic space and weight constraints, is currently well underway. 

\begin{figure}[htb!pb]
\centering
\includegraphics[width=0.98\columnwidth]{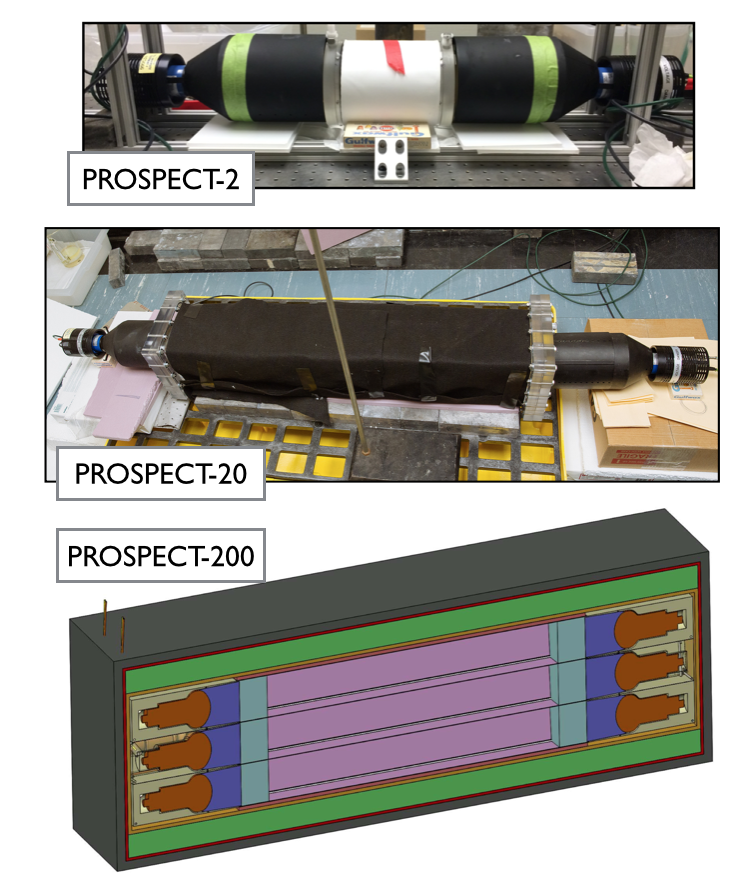}
\caption{PROSPECT prototype detector deployments/designs.  Top: PROSPECT2, Middle: PROSPECT20; Bottom: isometric drawing of the PROSPECT200 3x3 prototype.}
\label{fig:prototypes}
\end{figure}

Advances made during the R\&D process in the simulation, design, and production of passive shielding, a segmentation system, and liquid scintillator will be tested at the HFIR near site location in a suite of prototype PROSPECT detectors, as shown in Figure~\ref{fig:prototypes}.  A 1.7~liter cylindrical test cell with double-ended PMT readout, called PROSPECT2, has recently been assembled and tested at Yale and deployed at HFIR to take reactor-on and reactor-off data inside a passive shielding cave possessing similar qualities to that designed for the final PROSPECT experiment.  This prototype will be used to identify the dominant sources of background at the reactor site for the chosen PROSPECT detection method and demonstrate the background reduction capabilities of the PSD Li-loaded scintillator.  A single meter-long, 23~liter cell with cross-section similar to that of the baseline PROSPECT design, also possessing double-ended readout, called PROSPECT20, has also been assembled and tested at Yale for deployment at the HFIR near site in a similarly-designed shielding cave.  PROSPECT20 will be used to demonstrate PSD properties and position and energy reconstruction abilities in a full-length cell, while providing higher-statistics measurements of IBD-like signatures during reactor-on and reactor-off periods.  Finally, a prototype containing 9 full-length cells utilizing the exact design and fabrication techniques planned for the full PROSPECT experiment, called PROSPECT200, has been designed to demonstrate the feasibility of the existing PROSPECT design, and will also provide a demonstration of background reduction using topology-based cuts and target fiducialization.  Complete detector and background simulations of these prototypes are underway, and will provide the groundwork for benchmarking full PROSPECT simulations and producing credible background estimates prior to assembly and installation of the final detector.  Design of and results from PROSPECT2 and PROSPECT20 will be discussed in more detail in upcoming publications.

Deployment of detector prototypes in the intended HFIR near site location has also provided an opportunity to lay the groundwork for efficient deployment of the full PROSPECT detector.  PROSPECT collaborators, including HFIR personnel, are regularly on-site performing PROSPECT-related assembly or preparation work.  The collaboration now has significant experience working within the existing work plannning, work control, and certification regimes at HFIR and ORNL, and has an established procedure in place for working at HFIR.  Much experience has also been gained in methods of remote data transfer and run configuration and control at HFIR.

As with all precision experiments, calibration is an essential component of the development program. The oscillation analysis will require an excellent understanding of the relative efficiency and event rates between detector segments.  Techniques for the precise measurement of the volume of scintillator transferred to a detector are now well-established~\cite{dayabay_filling:2013}, and will be valuable in making relative comparisons between near and far detector. To determine target masses of individual detector cells, the collaboration is investigating methods of precision metrology of segment volumes prior to scintillator filling.  We must also determine the relative antineutrino detection efficiency of the detector segments from expected PSD, timing, and energy cuts.  Threshold effects can be controlled via a good understanding of relative energy scale. As demonstrated in recent oscillation experiments such as Daya Bay~\cite{DYB:2013cpc}, neutron captures and various alpha and gamma-rays emissions from intrinsic radioactive backgrounds can be utilized for this purpose, as can internally-deployed calibration sources.  Calibration deployment design and prototyping is currently underway, as are simulations of calibration regimes for the PROSPECT detector.

Measurement of the absolute antineutrino energy spectrum emitted by an HEU reactor will require additional precise calibration of the absolute energy scale. This must account for non-linear effects arising from escape of prompt energy from scintillating regions, light production in the scintillator, and processing of signals by the detector electronics.  The absolute energy scale calibration can be achieved using background or internal sources, as mentioned above. Beta emitters could be particularly useful in this instance, since the continuous energy spectrum can span a wide energy range and the comparison of measured and predicted shapes can be used to test detector response models.  Extensive bench-top characterization measurements of the scintillator, including measurements of the Birks parameters over a wide energy range, light absorption and re-emission characteristics, as well as dedicated measurements of electronics non-linearity, will also be required to develop such detector response models. In all of the cases mentioned above, simulation studies are being used to investigate the application of these techniques to a segmented scintillator detector.

\section{Conclusion}
\label{sec:conc}
The PROSPECT experiment, consisting of segmented scintillating detectors deployed at short baselines from a US-based research reactor, can provide precision measurements of the reactor antineutrino spectrum and probe anomalous electron neutrino disappearance results by searching for relative spectral distortion as a function of baseline.  A focused research and development program is underway to characterize potential experimental locations  and demonstrate that the required level of background rejection can be achieved. In addition to providing a definitive test of the ``reactor anomaly'', such an effort will provide a unique measurement of the $^{235}$U reactor antineutrino spectrum for use in improving reactor flux predictions. Furthermore, the detection technology developed to allow operation of antineutrino detectors near-surface will provide a revolutionary reactor safeguards capability,  enabling the deployment of monitoring detectors at a much broader range of locations for future non-proliferation efforts.

\bibliography{sblSnowmassWhitepaper}

\end{document}